\documentclass[12pt,preprint]{aastex}

\shorttitle{The eclipsing binary V69 in 47~Tuc}
\shortauthors{Thompson et al.}

\begin{document}

\title{The Clusters AgeS Experiment (CASE).~IV. \\
Analysis of the Eclipsing Binary V69 
in the Globular Cluster 47~Tuc\footnote{This paper includes data 
gathered with the 6.5-meter Magellan Baade and Clay  Telescopes and 
the 2.5-meter du Pont Telescope located at Las Campanas 
Observatory, Chile. It is based in part on data 
obtained at the South African Astronomical Observatory.}
}

\author{ 
I.~B.~Thompson\altaffilmark{2},
J.~Kaluzny\altaffilmark{3},
S.~M.~Rucinski\altaffilmark{4}, 
W.~Krzeminski\altaffilmark{5},
W.~Pych\altaffilmark{3},
A.~Dotter\altaffilmark{6},
G.~S.~Burley\altaffilmark{2} 
}

\altaffiltext{2}{Carnegie Observatories, 813 Santa Barbara St.,
Pasadena, CA 91101-1292; (ian,burley)@obs.carnegiescience.edu}

\altaffiltext{3}{Copernicus Astronomical Center, Bartycka 18,
00-716 Warsaw, Poland; (jka,pych)@camk.edu.pl}

\altaffiltext{4}{David Dunlap Observatory, Department of Astronomy and 
Astrophysics, University of Toronto, P.O. Box 360, Richmond Hill, 
ON L4C 4Y6, Canada; rucinski@astro.utoronto.ca}

\altaffiltext{5}{Las Campanas Observatory, Casilla 601, La Serena, 
Chile; wojtek@lco.cl}

\altaffiltext{6}{Department of Physics and Astronomy, University of Victoria, 
P.O. Box 3055, Victoria, B.C., V8W 3P6, Canada; dotter@uvic.ca}

\begin{abstract}
We use photometric and spectroscopic observations of the eclipsing
binary V69-47~Tuc to derive the masses, radii, and luminosities
of the component stars. Based on measured systemic velocity, 
distance, and proper motion, the system is a member of the globular 
cluster 47~Tuc. The system has an orbital period of $29.5~d$ and the
orbit is slightly eccentric with $e=0.056$. We obtain
$M_p=0.8762\pm 0.0048\,M_\odot$,
$R_p=1.3148\pm 0.0051\,R_\odot$, $L_p=1.94\pm 0.21\,L_\odot$
for the primary  and
$M_s=0.8588\pm 0.0060\,M_\odot$, $R_s=1.1616\pm 0.0062\,R_\odot$,
$L_s=1.53\pm 0.17\,L_\odot$ for the secondary. These components of 
V69 are the first Population~II stars with masses and radii derived
directly and with an accuracy of better than 1\%.
We measure an apparent distance modulus of $(m-M)_{V}=13.35\pm 0.08$ to V69.
We compare the absolute parameters of V69 with five sets  of stellar
evolution  models and estimate the age of V69 using mass-luminosity-age,
mass-radius-age, and turnoff mass - age relations. The masses, radii, and
luminosities of the component stars are determined well enough that the
measurement of ages is dominated by systematic differences 
between the evolutionary
models, in particular, the adopted helium abundance. By comparing
the observations to Dartmouth model isochrones we estimate the 
age of V69 to be 11.25$\pm0.21$(random)$\pm$0.85(systematic)~Gyr  
assuming [Fe/H]~=~-0.70, [$\alpha$/Fe]~=~0.4, and  $Y$~=~0.255. 
The determination of the distance to V69, 
and hence to 47~Tuc, can be further improved when infrared eclipse photometry
is obtained for the variable. 
\end{abstract}


\section{INTRODUCTION}
\label{intro}

Detached eclipsing double-line binary stars are the fundamental astrophysical
laboratory for the determination of stellar parameters of mass 
and radius. Luminosities can be derived using
measured parallaxes or from empirical color -- effective temperature
relations. These data are the fundamental tests of stellar
evolution models. Many field Population~I systems are known at solar mass 
and larger \citep{andersen} and modern high accuracy measurements
of masses, luminosities, and radii of the component stars are in general agreement 
with evolution models \citep[see, for example,][]{lacy05,lacy08,
clausen08}. Similar results are obtained for studies of individual
binaries in the old open clusters NGC~188 \citep{meibom09}, NGC~2243
\citep{kaluzny06}, and NGC~6791 \citep{grundahl08}. Recent wide field photometric
surveys have identified numerous low mass systems, and for these K and
M stars the common theme of a comparison of component properties with
evolution models
is that the models systematically underestimate the radii of the
components in these binaries.  Summaries of recent measurements
can be found in \citet{lopez06}, \citet{lopez07}, and \citet{blake08}. 

The situation is even less clear for Population~II stars. With the
exception of CM Dra, for which the component masses are 
$\sim$0.2$\,M_\odot$ \citep{lacy77}, and the $\omega$~Cen binary OGLEGC-V17, for
which the analysis is compromised by an uncertain determination of
the metallicity \citep{thompson01,kal02}, there are no known 
Population~II detached double-line eclipsing binaries with main-sequence
components.  \citet{torres02} used interferometric 
observations of HD~195987 ([Fe/H] $\sim$ -0.6) to derive an orbit and 
measure the masses of the components. The radiative properties of the 
two components agree with a suite of models given some slight
modifications of input parameters. While direct measurements of the radii 
of the components are not possible because HD~195987 is not an eclipsing
binary, estimates of the radii can be derived from the orbital parallax,
the bolometric flux,
and the estimated effective temperatures. Here again the measured radii 
are larger than the models by some ten per cent.
\citet{boyajian08} have measured the angular diameter of the G subdwarf
$\mu$~Cas ([Fe/H $\sim$ -0.8). This measurement provides a radius when
combined with the Hipparchos parallax for $\mu$~Cas. For this star the
models underpredict this measured radius by about five per cent. The 
masses of the components of $\mu$~Cas are only known to about ten per cent,
so the model comparisons here are not well constrained.

There is a clear need to locate and study Population~II detached
eclipsing binary stars to obtain accurate masses and radii of their
component stars. Stellar evolution models are becoming increasingly
sophisticated and are used to fit  observed cluster color-magnitude
diagrams (CMD). The cluster CMD's are themselves improving
in quality, with homogeneous surveys being conducted with the 
Hubble Space Telescope \citep[see, for example, ][]{sarajdeni07}.
Careful empirical tests of these models are an essential next step.

This is the first paper in a series devoted to the study of detached 
eclipsing double-line binaries (DEB) in Galactic globular clusters
with components on the cluster main sequence or subgiant branch. The 
Cluster AgeS Experiment (CASE) has the goal of determining
the basic stellar parameters (masses, luminosities, and radii) 
of the components of cluster binaries to a  
precision of better than  1\% in order to measure cluster ages 
and distances, and to test stellar evolution models. The methods and assumptions 
utilize basic and simple  approaches 
offered by the field of eclipsing double-line spectroscopic
binaries as described in \citet{paczynski97} and \citet{thompson01}.
Previous CASE papers have discussed blue straggler systems in 
$\omega$~Cen \citep{kal07a} and 47~Tuc \citep{kal07b}, and an SB1
binary in NGC~6397 \cite{kal08}.

The eclipsing binary  V69-47~Tuc (hereinafter V69) was discovered
by \citet{weldrake} during a survey for variable stars 
in the field of the globular cluster 47~Tuc. 
They presented an $I$-band light curve for the variable
and proposed an orbital period of $P=5.229$~d. The light curve phased 
with this period shows only one eclipse.  
The variable is located at the top of main sequence in the cluster
color-magnitude diagram, and thus is of potential great interest 
for measurements of the cluster age and distance. 

In this paper we report the  results of photometric and spectroscopic
observations aimed at a determination of the absolute parameters of 
the components of V69. 
Section~\ref{phot}
describes the photometry of the variable and the determination of an
orbital ephemeris. Section~\ref{spec} presents the
radial velocity observations. The combined photometric and 
spectroscopic element solutions are given in Section~\ref{comb}
while the membership in 47~Tuc is discussed in Section~\ref{memb}.
In Section~\ref{age} we compare the properties of the components 
of V69 to a selection of stellar evolution models with an emphasis 
on estimating the age of the system.
Finally in  Section~\ref{disc} we summarize our findings.

\section{PHOTOMETRIC OBSERVATIONS}
\label{phot}

The bulk of the photometric data were obtained with the 1.0-m Swope 
telescope at the Las Campanas Observatory using the 
$2048 \times 3150$ pixel SITE3 CCD camera 
with a scale of 0.435 arcsec/pixel.  
These observations were collected during the 2004 -- 2007 
observing seasons. The same set of $BV$ filters was used for all
observations. Exposure times ranged from 100~s to 360~s
for the $V$ filter (average exposure was 120~s) 
and from 160~s to 360~s for the $B$ filter (average exposure was 170~s).
An eclipse event was detected on the first of a total of 8 nights
of observations during 
the 2004 season. When the 2004 season data were combined
with the observations of \citet{weldrake} we were able to 
eliminate the proposed 5.25-d period, but the
combined data set was insufficient to establish a unique ephemeris. 
We then examined  the 286 $V$ band images collected 
by the OGLE team on 44 nights during the 1993 observing 
season for the cluster field 104-A \citep{kal98}. 
A single eclipse event was  detected on the night of 1993 July 22 UT 
and this permitted the identification of an approximate orbital period 
of $P\approx 29.540$~d.   
Subsequent observations collected in the 2005, 2006, and 2007 seasons  
concentrated on nights with predicted eclipse events. 
During the 2007 season we observed the variable 
with the 2.5-m du Pont telescope
using the $2048 \times 2048$ pixel TEK5 CCD camera at a scale 
of 0.259$\arcsec$/pixel. These data included one well covered primary
eclipse observed mostly in the $V$-band on 2007 August 18 UT.

In addition to the Las Campanas photometry, we obtained $V$-band 
observations of  an  eclipse on 2005 October 22 UT  using the 1.0-m telescope
at the South African Astronomical Observatory. Observations were obtained with
the STE4 $1024\times 1024$ pixel CCD camera at a scale of 0.31$\arcsec$/pixel.

\begin{figure}
\figurenum{1}
\label{fig1}
\plotone{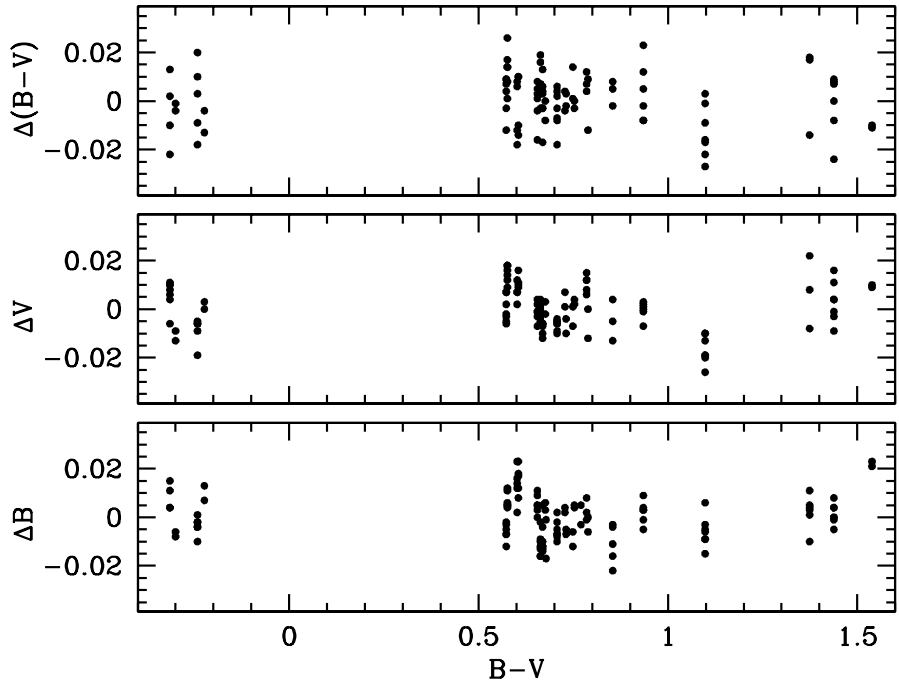}
\caption{
Plot of the color and magnitude residuals for the
standard stars observed on the night of 2007 August 19.
}
\end{figure}

Profile photometry was extracted for all of the observations
using the DAOPHOT/ALLSTAR package
\citep{stetson}. We calibrated  the instrumental photometry 
using observations collected with the du Pont telescope
on the night of 2007 August 19 UT. 
Observations were made of V69 together 
with  4 Landolt standard fields \citep{landolt92}. 
The standards were observed with a  range of air-mass of $1.19<X<1.94$.
The conditions were photometric 
and the seeing  ranged from 1.05 to 1.40 arcsec
with a median value of 1.22 arcsec.
The images of the cluster itself were obtained at an air-mass of 1.38 
with seeing of 1.0 arcsec. Profile photometry for the field of the 
variable was extracted from subframes covering $220\times 180$~arcsec
(the full field of TEK5 camera is $8.65\times8.65$~arcmin).
This helped to minimize the effects of a variable point-spread function
and as a result to obtain reliable aperture corrections. Aperture corrections 
for the V69 observations and the standard field observations
were derived using the  program DAOGROW \citep{stet90}. 
Magnitudes of 28 standard stars in the Landolt fields were taken from the Stetson 
catalog \citep{stet00}\footnote{We have used the electronic version of the 
catalog as of 2007 November 7. The catalog is maintained 
by Canadian Astronomy Data Centre at 
http://www3.cadc-ccda.hia-iha.nrc-cnrc.gc.ca/community/STETSON/standards/}. 
The following relations between instrumental (lower case letters)
and standard magnitudes were obtained:
\begin{eqnarray}
v=V-0.026(2)\times (B-V)+0.118(3)\times X + const, \\
b=B-0.069(2)\times (B-V)+0.217(3)\times X + const,
\end{eqnarray}
where $X$ is the air-mass. In Fig.~\ref{fig1}
we show the residuals between the standard and recovered
magnitudes for the Landolt primary  standards.
The analysed V69 field  includes 20 additional secondary standard stars from the 
Stetson catalog \citep{stet00}. The average residuals for these
stars are $\Delta V=+0.004 \pm 0.019$ and 
$\Delta (B-V)=-0.001 \pm 0.015$ with our magnitudes fainter and 
our colors bluer on average.  

\begin{figure}
\figurenum{2}
\plotone{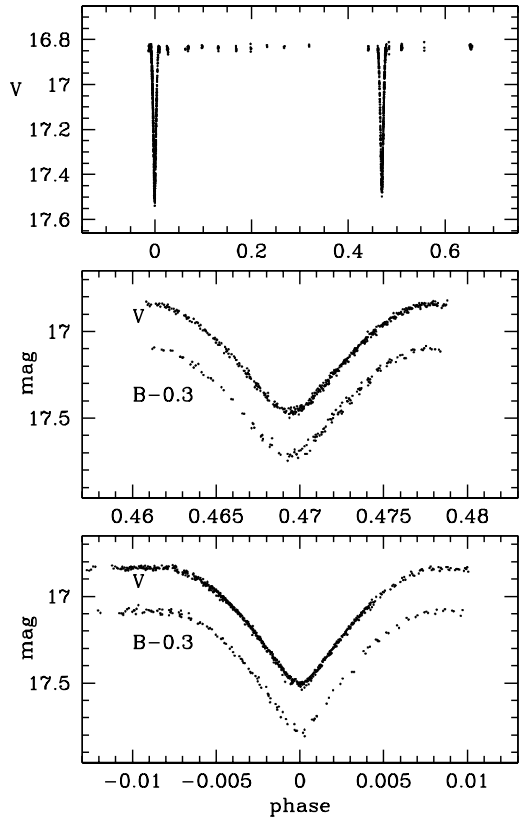}
\caption{
The phased $BV$ light curves of V69.
\label{fig2}}
\end{figure}

A total of 8 primary and 6 secondary eclipses were 
observed in the combined data sets.
In Fig.~\ref{fig2} we show the $BV$ light curves of V69 phased with 
the ephemeris :\\
\begin{equation}
Min I = HJD~245 3237.8421(2) + 29.53975(1)
\end{equation}
The period was derived using the Lafler-Kinman algorithm, and the error in
the period was estimated by visual inspection of the phased light curve 
as the period was varied away from the best fit value.
These light curves contain a total 
of 1216 and 310 data points for $V$ and $B$, respectively. The plots 
include photometry from all four data sets mentioned above.
For the OGLE data we have plotted only points inside the primary eclipse.
The colors and magnitudes of V69 at minima and at quadrature are listed 
in Table~\ref{tab1}. The quoted errors do not include possible systematic
errors of the zero points of the photometric solution which we estimate 
to be 0.010~mag.

\section{SPECTROSCOPIC OBSERVATIONS}
\label{spec}

We obtained spectroscopic observations of V69 with the  MIKE 
echelle spectrograph \citep{bern03}
on the Magellan Clay 6.5-m telescope.  All observations were taken with
a 0.7 arcsec slit at a resolution of R~$\simeq$~40,000.
The observations generally consisted of two exposures flanking an exposure of 
a thorium-argon hollow-cathode lamp. Total exposure times per
spectrum ranged from 1300 to 3665 seconds depending on 
observing conditions.
The observations were reduced with pipeline software written
by Dan Kelson following the approach of \citet{kel03}.
Post-extraction processing of the spectra
was done within the IRAF ECHELLE package.\footnote{IRAF 
is distributed by the National Optical Astronomy 
Observatories, which are operated by the Association of Universities
for Research in Astronomy, Inc., under cooperative agreement with the
NSF.} 

Velocities were measured with the TODCOR algorithm
\citep{zuc94} using an implementation written by G.\ Torres.
 For velocity templates
we used synthetic echelle resolution spectra
from the grid of \citet{coelho06}. We adopted
the values of log~g and $T_{eff}$ derived from the photometric solution
(see next section) and assumed a metallicity of ${\rm [Fe/H]} = -0.71$
with an $\alpha$-element enhancement of 0.4 (see the 
discussion of the metallicity of 47~Tuc in the next 
section). The templates were adjusted
once during the iterations for the Wilson-Devinney solutions, and final 
linear interpolations on the Coelho et al. grid were made at 
(log~$g$, $T_{eff}$ and [Fe/H]) 
= (4.14, 5945~K, $-0.71$) for the primary and 
(4.24, 5955~K, $-0.71$) for
the secondary. The measured velocities are insensitive to minor 
changes in these parameters. The interpolated synthetic spectra
were smoothed with a gaussian to match the resolution of the
observations. No rotational broadening was applied to the templates. 
The cross-correlations covered the wavelength intervals 
4125~\AA~$<$~$\lambda$~$<$~4320~\AA, 4350~\AA~$<$~$\lambda$~$<$~4600~\AA,
and 4600~\AA~$<$~$\lambda$~$<$~4850~\AA. The final adopted velocities
are the averages of these three measurements for each of the
observations.

\begin{figure}
\epsscale{1.0}
\figurenum{3}
\plotone{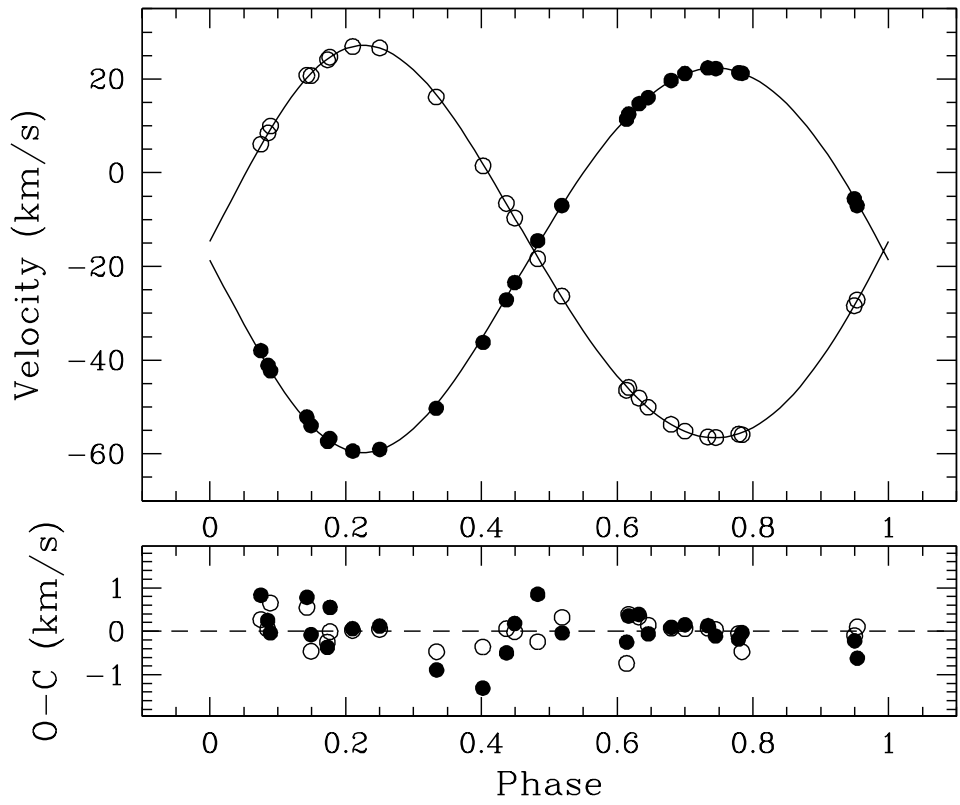}
\caption{
Radial velocity observations of V69. Filled symbols represent 
data for the primary and open symbols are for the secondary. 
\label{fig3}}
\end{figure}

The measured radial velocities were fit with 
a non-linear least squares solution
using code written by G.\ Torres adopting the ephemeris given in 
equation (3).
The observations are presented in Table~\ref{rvdata} which lists the 
heliocentric Julian Date (HJD) at mid-exposure, the velocities of the primary 
and secondary components, and the orbital phases of the observations. 
The adopted orbital elements are listed in Table~\ref{orbit} and 
the  orbit is plotted in Fig.~\ref{fig3}.

\section{LIGHT CURVE ANALYSIS AND SYSTEM PARAMETERS}
\label{comb}

We have used two different models for the analysis of the light curves. 
The first is the Wilson-Devinney model \citep{wd71,w79} as implemented in the
PHOEBE package \citep{prsa05}. The second is the 
JKTEBOB program \citep{sou04a,sou04b}, which is based on the EBOP code
\citep{popper80,PopperEtzel,Etzel81}. 
The most recent public version of JKTEBOB
is described in detail in \citet{sou07}. In particular, 
it incorporates an option to adopt a non-linear 
limb darkening law. 

The two light curve fitting programs utilize  different
methodologies and approaches. The PHOEBE/WD program is more 
sophisticated in its treatement of the geometry and the
stellar atmospheres. It utilizes Roche geometry to approximate the
shapes of the stars, uses  Kurucz model atmospheres, and 
treats the reflection effect in detail. It can be used for 
any component separation including contact systems.
JKTEBOP/EBOP approximates stars by bi-axial ellipsoids. It is faster
than the WD code but is appropriate only for binary components that
are spherical or only slightly distorted.
The real advantage of JKTEBOP is that the 
errors for the fit parameters are reliably determined. The
PHOEBE code provides only formal  errors of the fits and
these are known to be  underestimated (A. Prsa, personal
communication). However an advantage of the PHOEBE code is that it can be used
for the analysis of multi-color light curves. For
V69 the solution for the $B$-band is much 
poorer than for the $V$-band because of the lower quality 
of the $B$-band data. PHOEBE/WD  was used to 
simultaneously fit  the $V$ and $B$ curves. The resulting geometrical
parameters are based mostly on the higher quality $V$ curve while
the simultaneous fit permits a good estimate of the luminosity ratio. 

In the analysis we assumed that the light curve of V69 is 
free from any ``third light''. This is supported by the depths of the
eclipses, with $\Delta V>0.63$~mag for both the primary and secondary eclipse. 
There is also no evidence for 
any third component in the spectra of the binary. Finally, we note
that V69 can be located on numerous HST/ACS images (for example, PEP~ID~9018,  
P.I. G.~De~Marchi).
Examination of these images indicates that our ground-based photometry
does not suffer from any blending problems from unresolved visual 
companions to the variable.

Four recent high resolution studies of the metallicity of 47~Tuc 
suggest a value of ${\rm [Fe/H]} \sim -0.7$. 
As part of a study to establish a globular cluster metallicity
scale based on measurements of Fe~II lines, \citet{kraft03} 
reanalysed in a uniform way equivalent widths in a total
of 8 giants measured by \citet{brown92}, \citet{norris95}, 
and \citet{carretta97}. They obtained a range of -0.61 to -0.69  for 
[Fe/H]$_{\rm I}$  and a range of -0.56 to -0.70 for [Fe/H]$_{\rm II}$.
\citet{alves-brito}
used observations of 5 giants to obtain [Fe/H]$_{\rm I}$~=~-0.66
and [Fe/H]$_{\rm II}$~=~-0.69
with an $\alpha$-element enhancement of about 0.3 dex. 
\citet{koch} found [Fe/H]$_{\rm I}$~=~-0.76
and [Fe/H]$_{\rm II}$~=~-0.82 with an $\alpha$-element enhancement of 0.4 dex
based on a study of eight giants and one dwarf star. Finally,
\citet{carretta09} measured [Fe/H]$_{\rm I}$~=~-0.77 and 
[Fe/H]$_{\rm II}$~=~-0.80 with an $\alpha$-element enhancement of
+0.39 from observations of 11 giants.
For this analysis, we have adopted a value of
[Fe/H]~=~-0.71 together with an $\alpha$-element enhancement
of +0.4. Sensitivities of age estimates to these assumptions are 
explored in Sections 6.1 and 6.2.

The effective temperature of the primary is  needed for the
PHOEBE light curve solution and for both components to estimate the luminosities of the
stars from the derived radii. In the absence of detailed infrared
eclipse photometry we estimated the effective temperature of
the primary, $T_p$,  from the dereddened $(B-V)$  and $(V-I)$ color indices.
We used the observed 
color of the binary, $(B-V)$ = 0.548 (see Table~\ref{tab1}) together with
a reddening of $E(B-V)$ = 0.04  \citep{harris,gratton,percival02} to
derive $(B-V)_0$ = 0.508. For the $(V-I)$ color we note that V69
is star number 14065 in the catalog of \citet{kal98} where 
$(V-I)$ = 0.731$\pm$0.025. We shift this observed color by -0.026 to put
the observation on the system of \citet{stet00} \citep{percival02} and 
adopt the reddening law of \citet{schlegel} to obtain $E(V-I)$ = 0.06 and 
a final value of $(V-I)_0$ = 0.671.
The similarity of the $(B-V)$ colors in and out of eclipse 
(see Table~\ref{tab1}) suggests that the effective temperature of the primary
derived from these colors is close to the true value.

\begin{figure}
\epsscale{1.0}
\figurenum{4}
\plotone{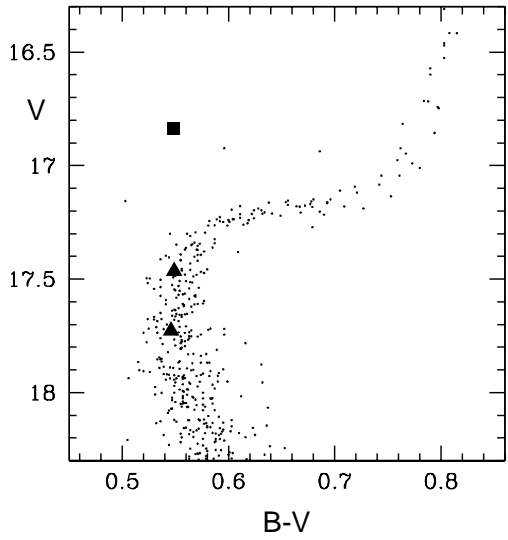}
\caption{
Position of V69 in the $BV$~CMD for
47~Tuc. The square gives the position for the combined
light and the triangles show the positions of each of the
individual components.
\label{fig4}}
\end{figure}

Visual inspection of the position of
the binary in a color-magnitude diagram suggests that the components
are beginning to evolve away from the main sequence (see Fig.~\ref{fig4}),
and so we adopt an initial gravity for the primary of $log~g~=~4.1$. 

These adopted colors and parameters lead to estimates for the effective temperature
of the primary of 6070~K and 5900~K, 6050~K and 5905~K, and 5960~K and 5780~K for
the $(B-V)$ and $(V-I)$ colors and the calibrations of \citet{worthey}, \citet{ramirez},
and \citet{vandenberg03}, respectively. We adopt a linear average of these
estimates, resulting in an effective temperature for the primary of
$5945\pm 105$~K, where the quoted error is the standard deviation of the six temperature
estimates.

The three color-temperature relations considered here do not explicitly account for
$\alpha$-element enhancement.  Based on the synthetic color-temperature employed by
\citet{dotter}, we estimate that the difference in $T_{eff}$ between [$\alpha$/Fe]=0
and +0.4 is $\sim$10~K for the [Fe/H], $T_{eff}$, and log~g of the primary.  We
expect this is an upper limit because the color-temperature relations should include
some implicit dependence on [$\alpha$/Fe].  In any case, the [$\alpha$/Fe]-color
uncertainty is small compared to that which arises between the different
color-temperature relations or between $B-V$ and $V-I$.

We note that \citet{koch} adopted an effective temperature of 5750~K 
for the turnoff star that they studied based on excitation equilibrium 
for Fe~I lines, and it is of interest to ask how accurate our photometric
estimate of the effective temperature of the primary of V69 actually is.
We estimate the uncertainty in the reddening to be 0.010~mag, and
this value coupled with our estimated uncertainty in the photometric
calibration suggests that the uncertainty in the derived unreddened
$(B-V)$ colors of the components of V69 to be $\sim$0.014~mag. The 
$(V-I)$ color is somewhat less accurate.
For a range of metallicity of [Fe/H] from $-0.61$ to $-0.77$ (see above),
an estimated range in the gravity of 0.2~dex and a  uncertainty in $(B-V)_0$ of
0.014, the Worthey-Lee calibration suggests an uncertainty in the derived
effective temperature of approximately 80~K.  The calibrations of \citet{ramirez},
and \citet{vandenberg03} indicate similar uncertainties. Considering the 
scatter about $T_{eff}$ in the 
various calibrations we conservatively conclude that we can estimate the 
effective temperature of the components of V69 to an accuracy of about 150~K.
These three 
calibrations predict bolometric corrections spanning the range  $-0.11$
to $-0.09$ and we  adopt $BC_{p}=BC_{s}=-0.10\pm 0.01$.

\subsection{\it Models Using the PHOEBE Code}

The PHOEBE implementation of the Wilson-Devinney  model fits 
the orbital inclination $i$, 
the gravitational potentials $\Omega_{p}$ and
$\Omega_{s}$, the effective temperature of the secondary $T_{s}$,
the eccentricity $e$, the longitude of the
periastron $\omega$ and the
relative luminosities $L_s$/$L_p$ in $B$ and $V$ simultaneously.
We iterated on the solution, calculating 
subsequent values for the $(B-V)_0$ colors for the primary and
secondary from the observed $(B-V)_0$ for the system and the
modeled values for $(L_s/L_p)_B$ and $(L_s/L_p)_V$ at each
iteration. The mass ratio was set to the spectroscopic value
of $q~=~mass_s/mass_p~=~0.984$. The iterations converged in
two cycles.
The final values of the fit
are listed in Table~\ref{lc2}. The values for the relative radii
are derived  directly from the non-dimensional potentials $\Omega_{1}$ and
$\Omega_{2}$, and the spectroscopic mass ratio $q$. The luminosity
ratios and the relative radii are insensitive to changes in effective temperature:
$(L_s/L_p)_B$, $(L_s/L_p)_V$, $r_s$, and $r_p$ all change by less than
0.3 per cent for a $\pm$150~K change in effective temperature.

\subsection{\it Models Using the JKTEBOB Code}
   
The JKTEBOB code optimizes the sum of the relative radii $r_{p}+r_{s}$,
the ratio $k=r_{s}/r_{p}$, the ratio of the surface brightness 
of the two stars $J$, the orbital eccentricity $e$, and the longitude of the 
periastron $\omega$. These two last parameters were included in the
analysis by fitting 
$e$~cos~$\omega$ and $e$~sin~$\omega$. The mass ratio was fixed at 
the spectroscopic value of $q = 0.984$, 
and the gravity brightening exponent was set
to 0.32. Note that the values for these two parameters have a negligible    
effect on the model light curve as the stars are very well separated and
both are practically spherical.  We adopted  a square-root law for the 
limb darkening and adopted theoretical limb darkening  coefficients 
from \citet{van93} for $T_{eff}=5959~K$ (adopted from the PHOEBE
code solution)  and ${\rm [Fe/H]=-0.71}$.  

\begin{figure}
\epsscale{1.0}
\figurenum{5}
\plotone{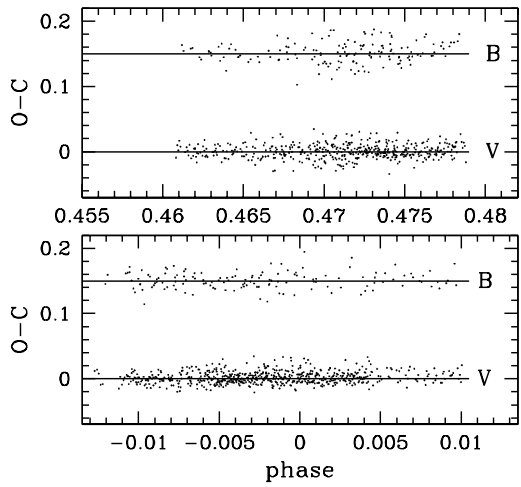}
\caption{
The residuals of the fits to the light curve  obtained with JKTEBOP
for the primary (bottom) and secondary eclipse (top).
The residuals for $B$ are offset by 0.15 mag for clarity.
\label{fig5}}
\end{figure}

Table~\ref{lc1} contains the values of the fit parameters along
with other relevant parameters derived from the solution. 
The last two rows list the reduced chi-squared of the fit
and the rms of the residuals. Separate solutions were 
derived for light curves in the $V$ and $B$ bands. The uncertainties of 
individual parameters were estimated using Monte Carlo simulations
as described in \citet{sou04b}. Ten thousand simulations were run 
for each light curve.  The solution based on the $V$ light
curve is much better constrained than the solution  for the $B$ band. 
The residuals of the fit obtained with 
the JKTEBOB are shown in Fig.~\ref{fig5}.

\subsection{\it Adopted Stellar Parameters}
\label{adopted}

A comparison of Tables~\ref{lc2} and \ref{lc1} shows that the two
light curve synthesis codes give very similar results.  
There is also  good agreement between the 
fitted values of $e$ and $\omega$ and those 
derived from the spectroscopic observations.
The remarkably small errors for the fit radii and relative luminosities
might seem suspicious at 
first given that the light curves suggest that the
eclipses of V69 are partial. However, the  
secondary eclipse is very close to being total. Examination of the
synthetic light curves shows that only  0.4\% of the 
surface of the secondary component remains visible at the center 
of the secondary eclipse. 

The magnitudes for the individual components
of the binary are derived from the luminosity ratios from the light
curve solution and the observed out-of-eclipse photometry. We obtain
$V_{p}=17.468\pm 0.010$, $B_{p}=18.019\pm 0.010$,
$V_{s}=17.724\pm 0.010$ and  $B_{s}=18.268\pm0.010$ where the errors 
are dominated by the contribution from zero point uncertainties of 
our photometry, estimated  at 0.010 mag. The colors of both components are
practically identical with $(B-V)_{p}=0.551\pm0.014$ and 
$(B-V)_{s}=0.544\pm 0.014$. The position of the binary on the 
47~Tuc color -- magnitude diagram is shown in Fig.~\ref{fig4}. 
The primary component has left the main-sequence of 
the cluster and is starting to ascend onto the subgiant branch. The secondary
also has begun to evolve and is among the bluest stars at the
cluster turnoff.

The absolute parameters of V69 obtained from our spectroscopic and
photometric analysis are given in Table~\ref{elements}. Because
the JKTEBOP code provides  better error estimates than PHOEBE
we have adopted the $V$-band  relative radii from Table~\ref{lc1}. 
Note, however, that the relative radii derived 
with the PHOEBE models agree at a level of 0.2\% with those obtained with 
the JKTEBOP models. The luminosities of the components follow
directly from
$log(L/L_\odot) = 2~log(R/R_\odot) + 4~log(T_{eff}/T_\odot)$
where  we have adopted $R_\odot$ = 6.9598~$\times$~$10^{5}$~km \citep{bahcall05}, 
and $T_\odot$ = 5777K \citep{neckel86}.

\section{MEMBERSHIP OF V69 AND THE DISTANCE OF 47 TUC}
\label{memb}

Before using the derived stellar parameters of the components of
V69 to estimate the age of 47~Tuc, it is appropriate to consider the
cluster membership of the star. 
There are four lines of evidence available.
First, although the velocity of 47~Tuc is low at 
$v_{rad}=-18.7\pm 0.5~km~sec^{-1}$  \citep{gebhardt95}, and thus does
not provide a strong discriminant against projected field stars,
the center-of-mass velocity of V69 ($\gamma = -16.36~km~sec^{-1}$)
is consistent with this velocity. At the location of the variable --
about 6 arcmin from the cluster center -- the velocity dispersion of
cluster stars is about 8~km~sec$^{-1}$.  
Second, Fig.~\ref{fig4} shows that the individual components of V69
lie on the cluster main sequence. 
Third, the binary is a proper motion member of 47~Tuc. 
The absolute proper motion of 47 Tuc has been measured to be 
$\mu_{\alpha}~=~5.64\pm$  0.20 mas/yr and $\mu_{\delta}~=~-2.05\pm$ 0.20 mas/yr 
\citep{anderson}.  V69 is located in Field F of that study. 
This field has subsequently become a standard calibration field for HST,
and V69 is present in numerous ACS/WFC frames taken between 2002 and 2006.
A quick study of the $\sim$30s F606W exposures shows that the proper
motion of V69 relative to the bulk of the cluster stars is 0.2 mas/yr in RA 
and -0.1 mas/yr in Dec, a typical internal motion for a cluster
member at this radius (J. Anderson, personal communication).  
Finally, based on the absolute magnitudes listed in 
Table~\ref{elements}
and the apparent magnitudes listed in Sec.~\ref{adopted}
we  derive distance moduli for the components of V69 of 
$(m-M)_{V}=13.35\pm 0.11$ and $(m-M)_{V}=13.34\pm 0.12$ for the primary and 
secondary, respectively. The average of these values is
$(m-M)_{V}=13.35\pm 0.08$. We compare this value  to some 
recent  distance determinations for 47~Tuc in Table~\ref{distance}.
While \citet{bono} have shown that there appears to be some systematic 
errors in the measurement of the distance to 47~Tuc by different techniques, V69 clearly
lies at the distance of 47~Tuc. 
We note in particular that \citet{kal07b} find a distance modulus of 
$(m-M)_{V}=13.40\pm 0.08$ for OGLE-228, another eclipsing binary 
in 47~Tuc. We conclude that V69 is a member of the globular cluster 47~Tuc.

\section{THE AGE OF 47 TUC}
\label{age}

The potential of using observations of eclipsing binaries for a robust 
determination of globular cluster ages was advocated by \citet{paczynski97}.
In particular, he argued that the mass-luminosity-age relation be used
rather than the mass-radius-age relation since the latter can be affected
by inaccuracies in the stellar models related to the treatment of 
sub-photospheric convection. In the following we compare the properties of 
the components of V69 with five different sets of stellar evolution models. 
We have attempted to minimize the effects of metallicity and $\alpha$-element
enhancement 
by comparing models with uniform values of [Fe/H] = -0.71 and, where possible, 
${\rm [\alpha/Fe}]$ = +0.4.

In our analysis we use the Dartmouth Stellar Evolution Database \citep{dotter}, 
the Padova models \citep{salasnich00}, the Teramo models \citep{piet06}, the 
Victoria-Regina models \citep{vandenberg}, and the Yonsei-Yale models 
\citep{kim02}.\footnote{The models and supporting software are available at 
the following websites: 
Dartmouth (http://stellar.dartmouth.edu/$\sim$models/), Padova (http://pleiadi.oapd.inaf.it/),
Teramo (http://193.204.1.62/index.html), Victoria-Regina 
(http:www.cadc-ccda.hia-iha.nrccnrc.gc.ca/cvo/community/VictoriaReginaModels/),
and Yonsei-Yale (http://www.astro.yale.edu/demarque/yyiso.html).}
Table~\ref{models} compares the parameters of each model set used in the 
derivation of the ages of V69. Entries for the solar abundance scale refer to
\citet{grevesse93} (Z/X~$\sim$~0.024)) and \citet{grevesse98} 
(Z/X~$\sim$~0.023). We note that solar models based on recent 
measurements of the solar abundance scale remain in conflict with helioseismology data
(see \citet{serenelli09} for a discussion).
We compare the derived values of mass, luminosity, 
and radius for the components of V69 with the isochrones. For the Dartmouth,
Victoria-Regina, and Yonsei-Yale models we used interpolation software provided 
by the different groups  to generate the plotted isochrones. For the Padova models 
we linearly interpolated on log~(age) to generate isochrones spaced by 0.5~Gyr for 
ease of comparison with the other models.

\begin{figure}
\epsscale{1.0}
\figurenum{6}
\plotone{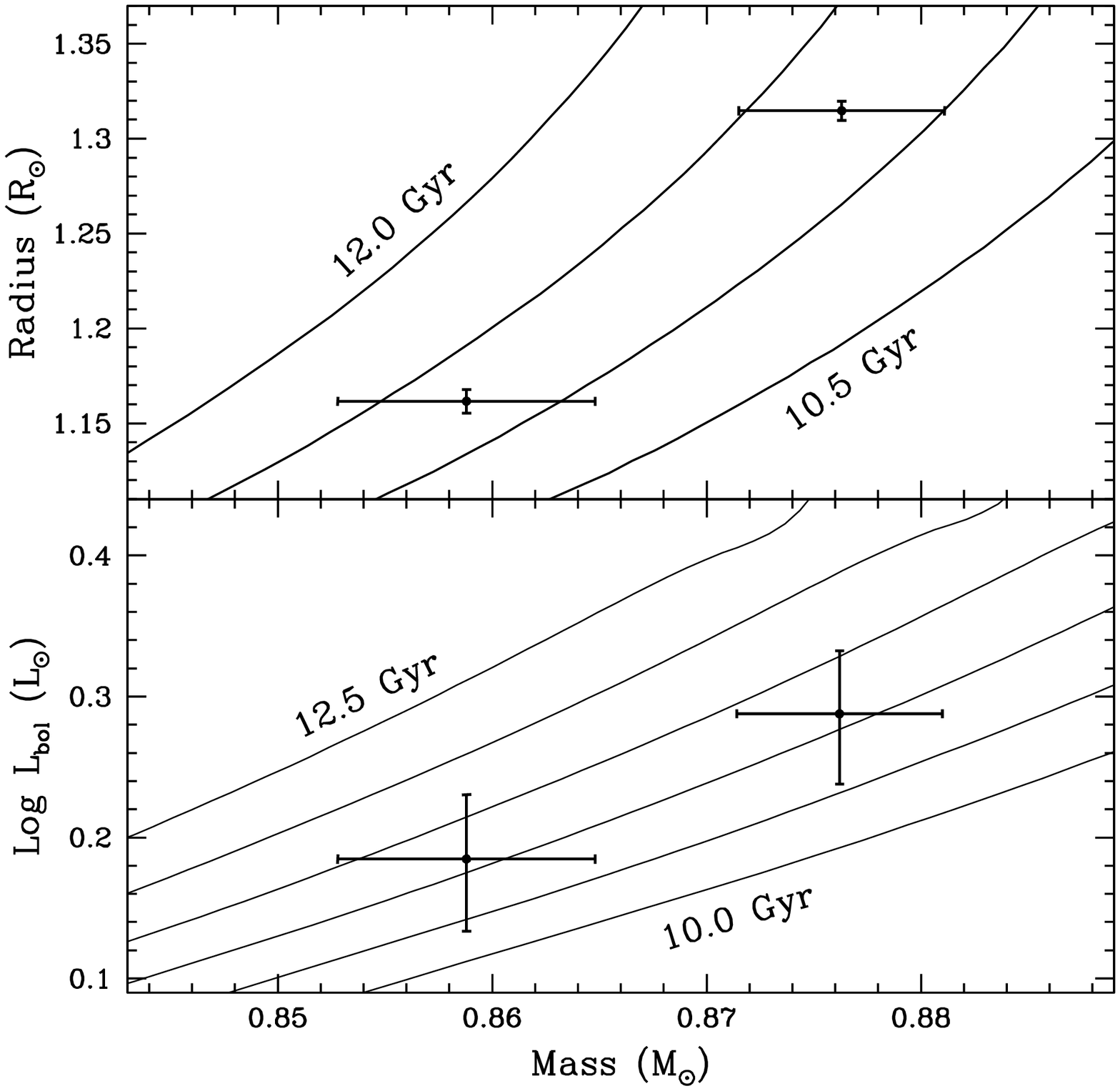}
\caption{
The masses, luminosities, and radii  of the components of V69 are compared to 
isochrones based on the Dartmouth models. For each panel the isochrones
are plotted in steps of 0.5~Gyr, with the lowest and highest age
isochrones labeled.
\label{fig6}}
\end{figure}

\begin{figure}
\epsscale{1.0}
\figurenum{7}
\plotone{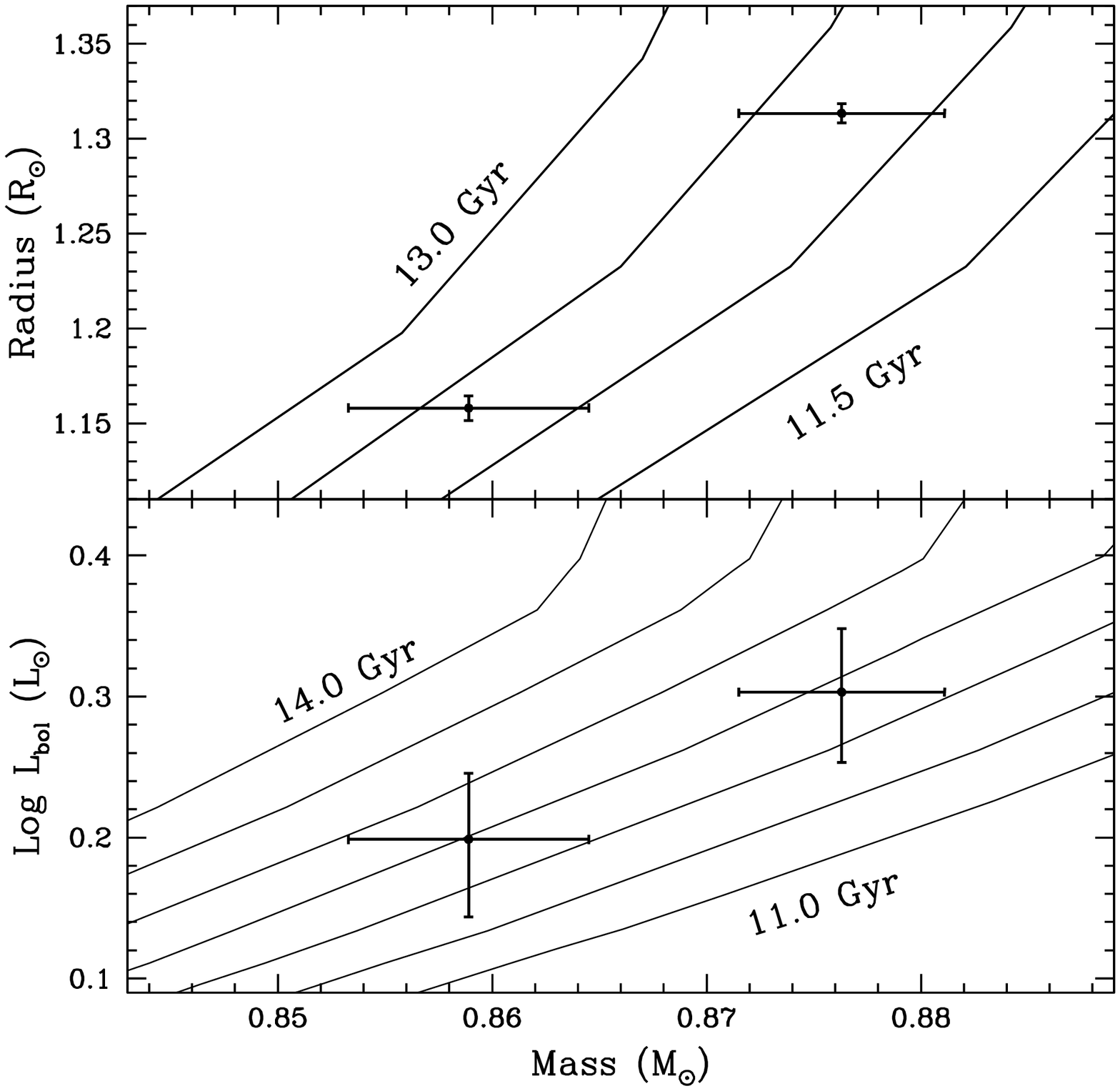}
\caption{
The masses, luminosities, and radii  of the components of V69 are compared to 
isochrones based on the Padova models. For each panel the isochrones
are plotted in steps of 0.5~Gyr, with the lowest and highest age
isochrones labeled.
\label{fig7}}
\end{figure}

\begin{figure}
\epsscale{1.0}
\figurenum{8}
\plotone{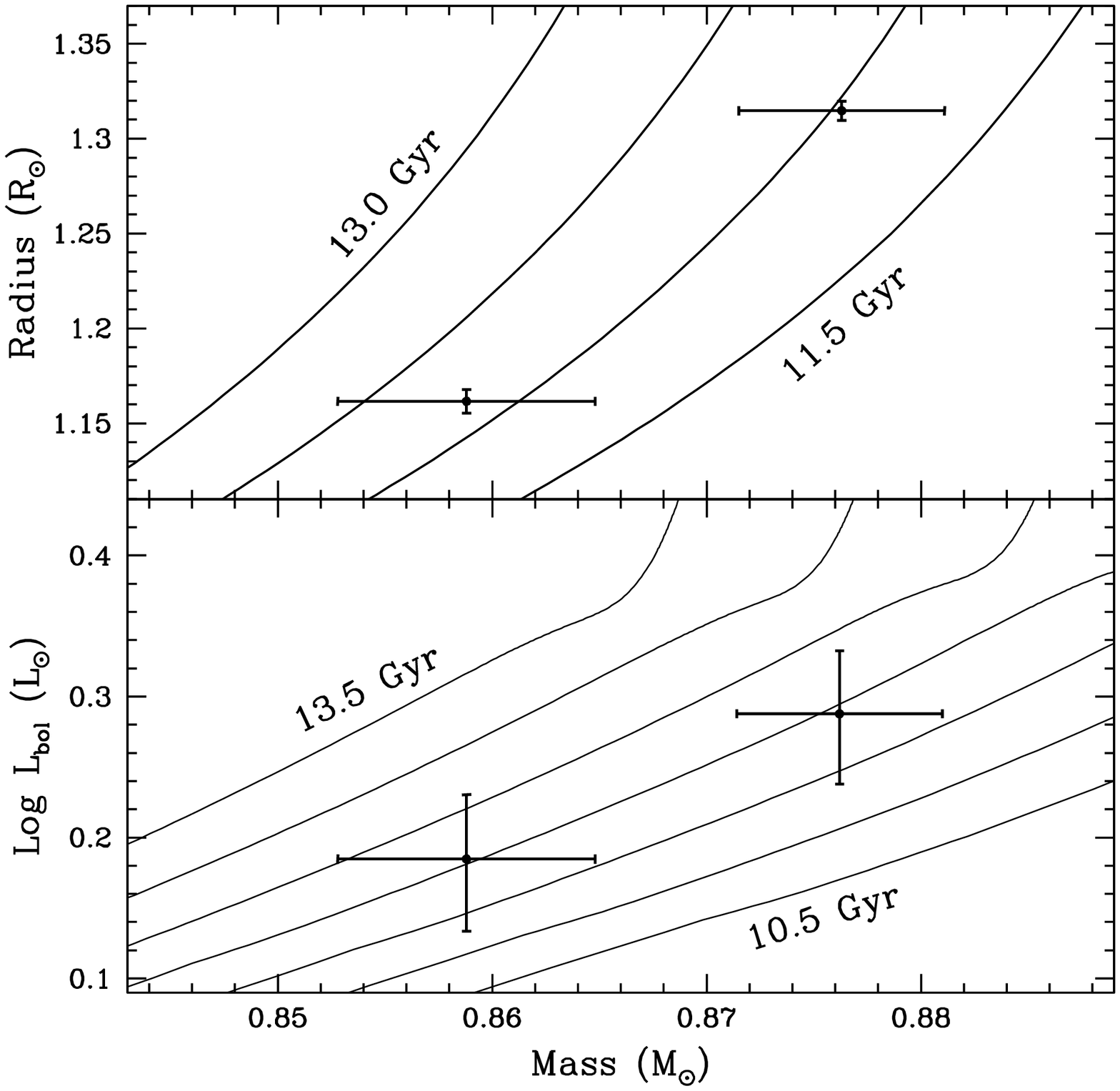}
\caption{
The masses, luminosities and radii of the components of V69 are compared to 
isochrones based on the Teramo models. For each panel the isochrones
are plotted in steps of 0.5~Gyr, with the lowest and highest age
isochrones labeled.
\label{fig8}}
\end{figure}

\begin{figure}
\epsscale{1.0}
\figurenum{9}
\plotone{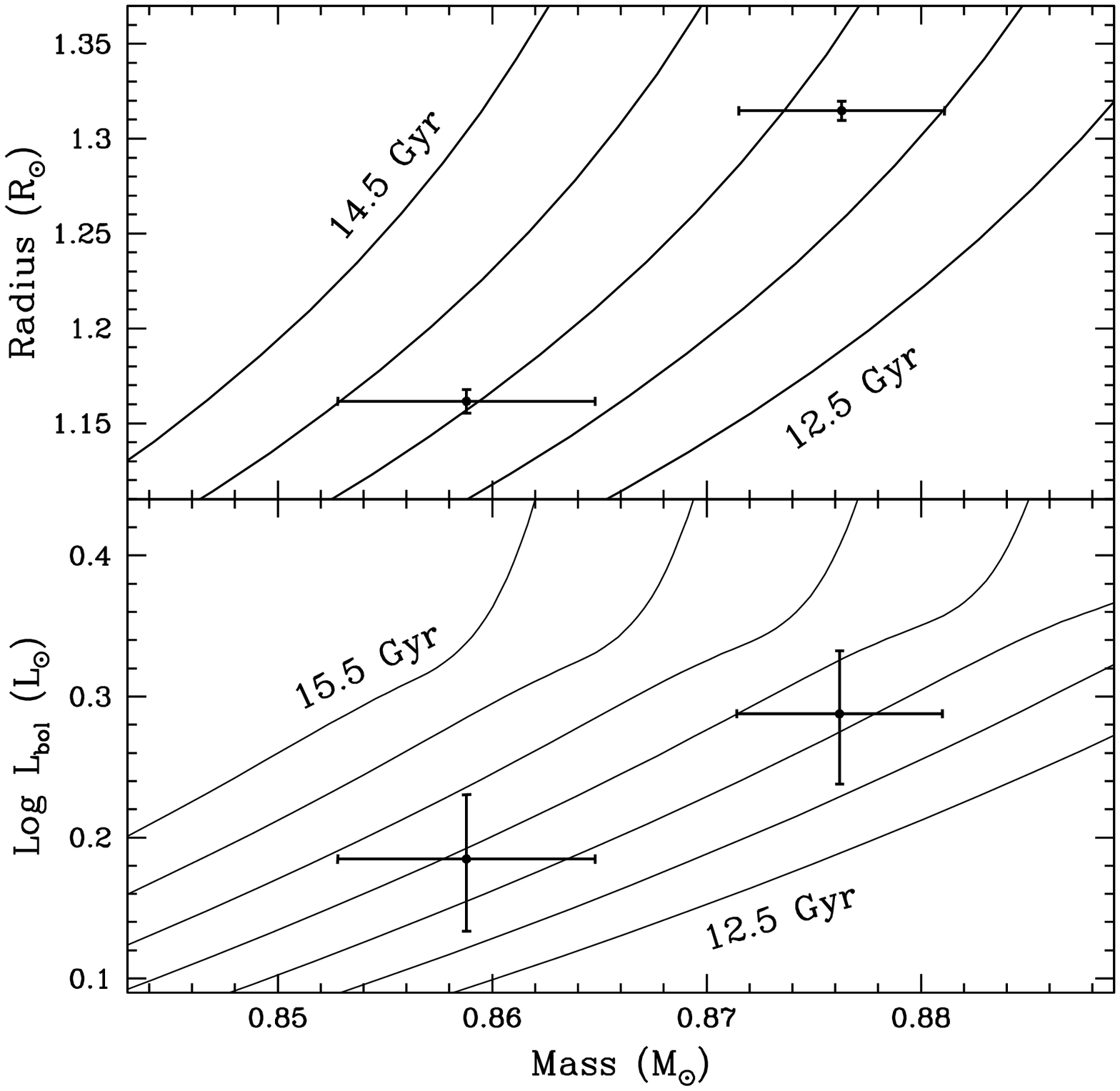}
\caption{
The masses, luminosities, and radii of the components of V69 are compared to 
isochrones based on the Victoria-Regina models. For each panel the isochrones
are plotted in steps of 0.5~Gyr, with the lowest and highest age
isochrones labeled.
\label{fig9}}
\end{figure}

\begin{figure}
\epsscale{1.0}
\figurenum{10}
\plotone{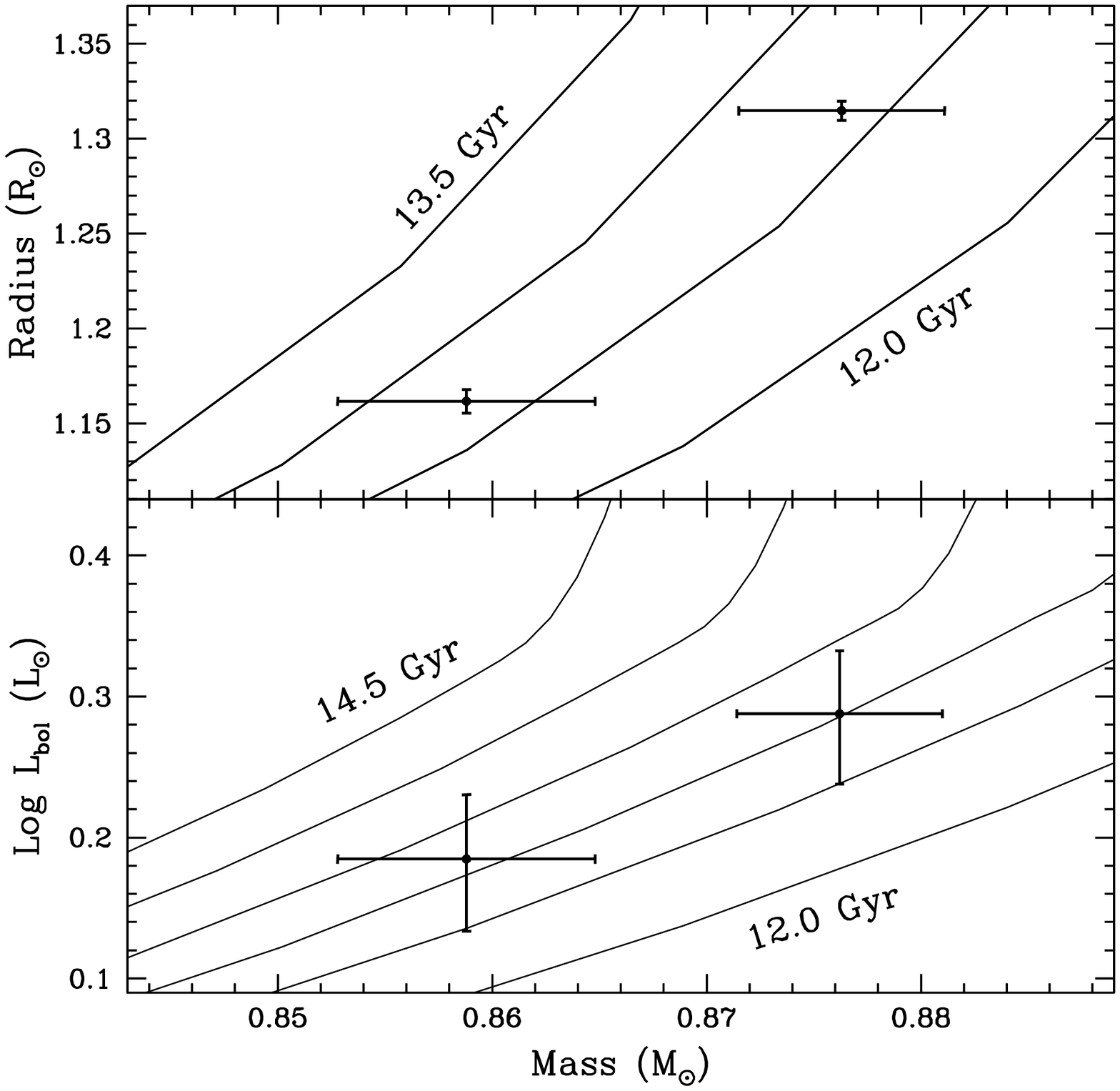}
\caption{
The masses, luminosities, and radii of the components of V69 are compared to 
isochrones based on the Yonsei-Yale models. For each panel the isochrones
are plotted in steps of 0.5~Gyr, with the lowest and highest age
isochrones labeled.
\label{fig10}}
\end{figure}

We plot the components of V69 on the mass-luminosity and mass-radius planes for the  
Dartmouth, Padova, Teramo, Victoria-Regina, and Yonsei-Yale models in 
Fig.~\ref{fig6}, Fig.~\ref{fig7}, Fig.~\ref{fig8}, Fig.~\ref{fig9}, and Fig.~\ref{fig10}, 
respectively. In each Figure isochrones are shown with a step of 0.5~Gyr.    
The measured ages for the components of V69 are summarized in Table~\ref{ages}. 
To calculate the random errors in the age estimates we assumed that the measurement 
errors in mass, luminosity and radius are uncorrelated. This is a reasonable assumption 
for the errors in luminosity are dominated by the adopted error in $T_{eff}$ and the 
errors in radius are dominated by the photometric solution rather than the error in 
$A~sin~i$ as measured by the orbital solution. The errors in mass, luminosity, and 
radius lead to uncertainties in age through comparison with the sets of isochrones, 
and the final random errors listed in Table~\ref{ages} are the quadrature sums of the 
mass-radius and mass-luminosity age uncertainties. The uncertainties in the masses 
(about 0.6\%) and  in the luminosities (about 10\%) give similar contributions to the 
errors of the derived ages, while the random errors in the age estimated from mass-radius-age 
relations are completely dominated by the measurement errors in the masses of the
components.  We emphasize that the plotted uncertainty in luminosity arises directly
from the $\pm$150~K uncertainty in the effective temperatures of the components of V69.
The radii are measured with a better relative accuracy than the luminosities and therefore 
the ages can be derived with a random error a factor of two smaller than the ages derived 
from the mass-luminosity relations. Table~\ref{ages} also presents the weighted average 
of the ages measured for the primary and secondary components of the binary.

\begin{figure}
\epsscale{1.0}
\figurenum{11}
\plotone{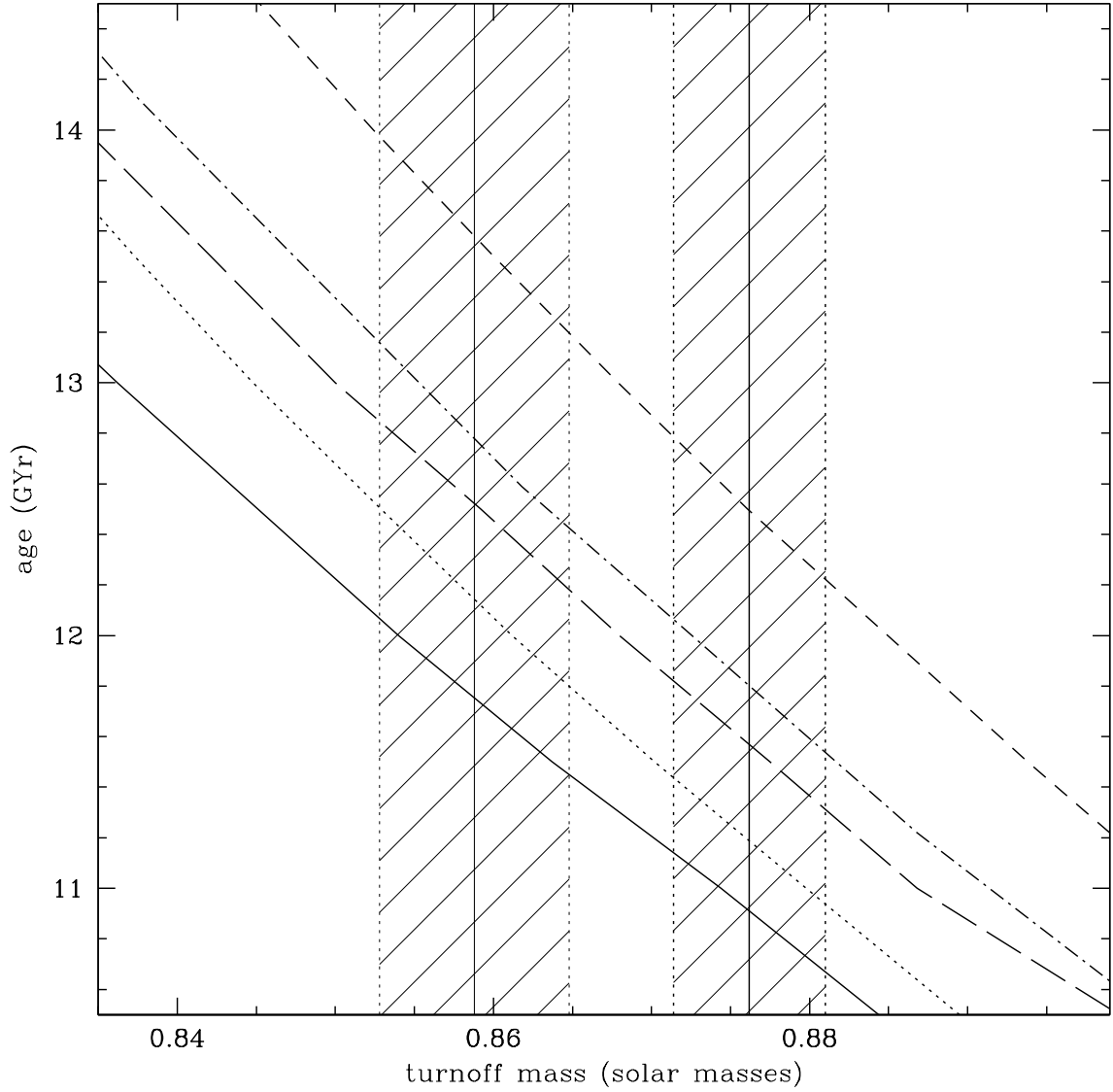}
\caption{
Age as a function of turnoff mass for the Dartmouth
(solid line), Teramo (dotted line), Yonsei-Yale (long dashed line),
Padova (dot-dashed line),
and Victoria-Regina (short dashed line) models. The masses for the
primary and secondary components of V69 are plotted as solid
vertical lines, with one-sigma errors represented by the shaded  
areas about the component masses. 
\label{fig11}}
\end{figure}

Finally, since the components are very close to the cluster turnoff (see Fig.~\ref{fig4}), 
the age of the system can be estimated from turnoff mass-age relations. For each isochrone 
we estimated the turnoff mass as the mass at maximum $log~T_{eff}$ in that isochrone. 
Based on the position of the components in the CMD we conclude that the secondary is at 
the blue extreme of the CMD, and that the primary has begun to evolve onto the subgiant 
branch at a slightly cooler effective temperature. Our mass measurements are accurate 
enough to resolve the progression of mass along an isochrone in this part of the CMD. 
Fig.~\ref{fig11} shows the resulting relations, and these derived ages are also summarized 
in Table~\ref{ages}. For completeness, ages are also given for the primary 
assuming it is at the cluster turnoff. The quoted errors in these ages 
are derived assuming only an error in the measurement of the masses, and 
are measured from the values of the ages at the $\pm1 \sigma$ extrema of the 
measured masses. This age measurement is obviously fairly crude, but serves
as a consistency check on the other ages estimates.

The ages of V69 determined from these relations show some considerable spread from one model set to the next, ranging from 11.1 to 13.7~Gyr. While we only have age estimates 
from two stars, we can draw some general 
conclusions. The first is that the close agreement in the age estimates for the primary 
and secondary for all models suggests that the errors in luminosity are overestimated. 
Second, the ages measured by the three methods also agree to within the random errors 
within each model. This 
suggests that concerns about the model radii arising from uncertainties in the treatment of
convection are overstated, and that the models predict internally consistent luminosities 
and radii at the measured masses. Future studies of clusters with more binaries covering 
a wider range of mass will be important in addressing this second issue.

We next investigate whether or not there are systematic differences in the different model 
sets that might explain the wide range in measured ages. Many variables will affect the 
isochrones, among the most significant are adopted $\alpha$-element enhancement and the 
relative abundance distribution over the $\alpha$-elements, helium abundance, 
mixing length and convective overshoot, and the 
treatment of diffusion and gravitational settling. None of the models considered include the
effects of rotation.  Two of these variables ([Fe/H] and $\alpha$-element enhancement) 
are measurable 
from spectroscopic data, while the others are parameters of the models.

In the following subsections we explore the sensitivity of the measured ages to [Fe/H],
[$\alpha$/Fe], diffusion and gravitational settling, and helium abundance.

\subsection{\it Sensitivity to Adopted [Fe/H]}

\begin{figure}
\epsscale{1.0}
\figurenum{12}
\plotone{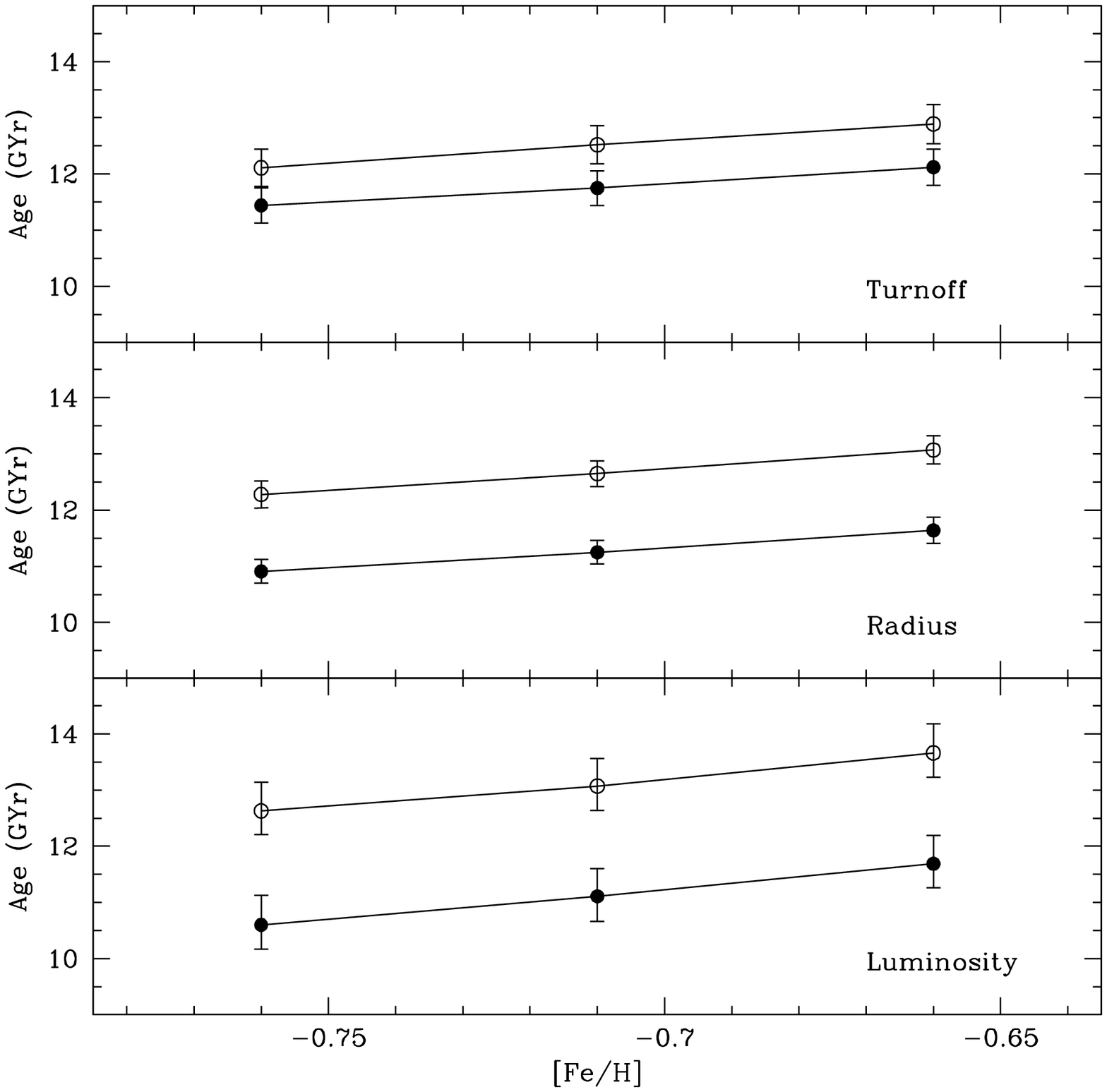}
\caption{
Age as a function of [Fe/H] for Dartmouth (filled 
symbols) and Yonsei-Yale (open symbols) models. Ages are determined from 
mass-luminosity relations (bottom panel), mass-radius relations (middle
panel), and turnoff mass - age relations (upper panel).
\label{fig12}}
\end{figure}

Figure~\ref{fig12} shows the sensitivity of the measured ages to model [Fe/H] for
the Dartmouth and Yonsei--Yale models sets for a range of $\pm$0.05 dex 
around our adopted value of -0.71 for each of the three methods of measuring
age. These values are interpolated in the models at a fixed [$\alpha$/Fe]~=~+0.4.
The plotted values are a weighted average of ages measured for the primary and secondary.
The slopes (in terms of $\Delta$~Gyr per 0.1 dex) are given in 
Table~\ref{alpha-fe}. In all three cases the slopes for the two model sets
are very similar, and Table~\ref{alpha-fe} also gives the average slopes.

\subsection{\it Sensitivity to $\alpha$-Element Enhancement}

\begin{figure}
\epsscale{1.0}
\figurenum{13}
\plotone{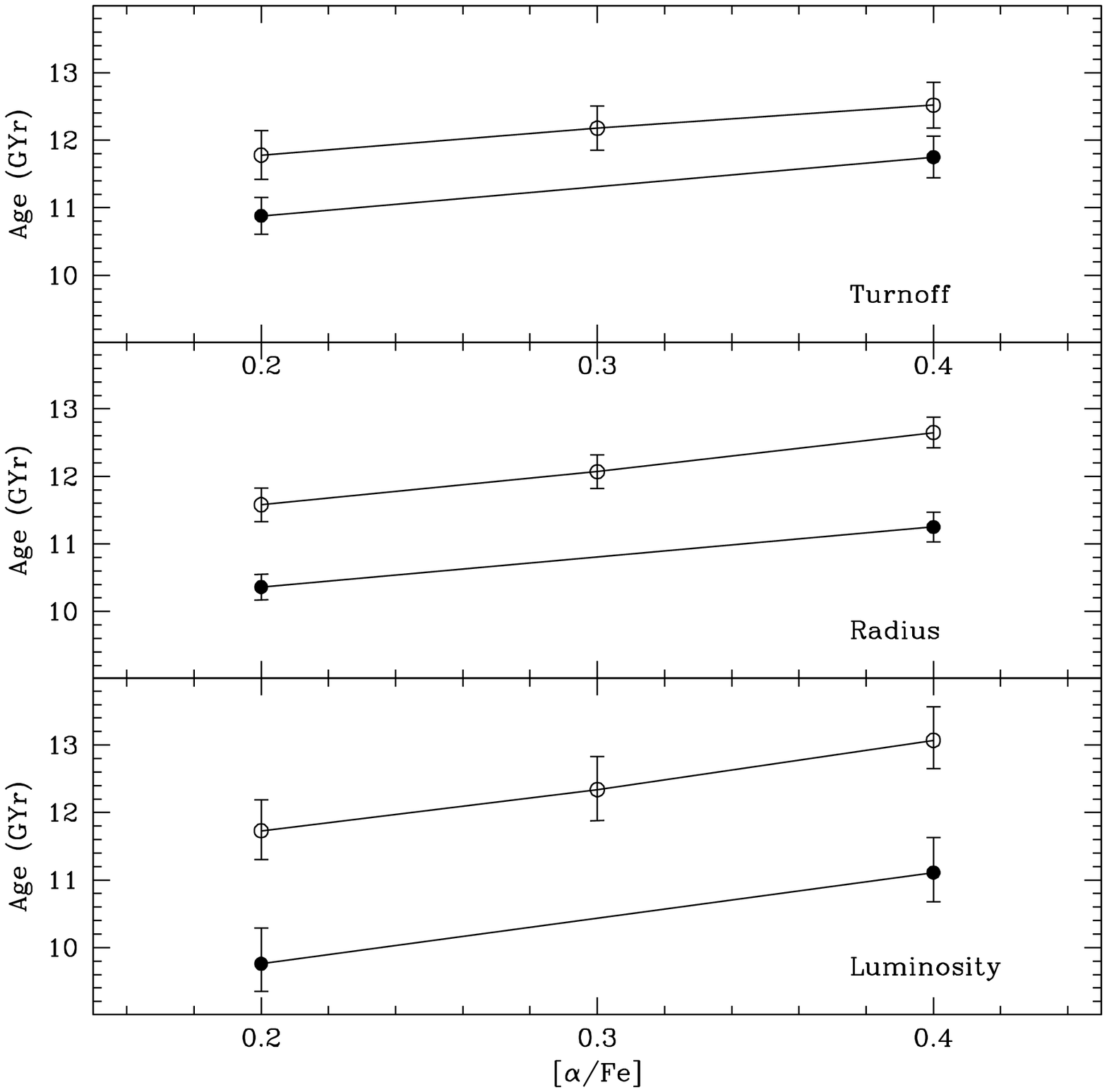}
\caption{
Age as a function of $\alpha$-element enhancement for Dartmouth (filled 
symbols) and Yonsei-Yale (open symbols) models. Ages are determined from 
mass-luminosity relations (bottom panel), mass-radius relations (middle
panel), and turnoff mass - age relations (upper panel).
\label{fig13}}
\end{figure}

We have estimated the effect of differing $\alpha$-element enhacement by
measuring the ages of the components of V69 from Dartmouth and Yonsei-Yale
isochrones calculated to have differing $\alpha$-enhancement in the range 0.2 to
0.4 at a fixed [Fe/H] = -0.71. The results are shown in Fig.~\ref{fig13},
where we plot the weighted average of the ages measured for the primary and
secondary components against model [$\alpha$/Fe]. As we found for the
sensitivity to adopted model [Fe/H], the slopes (in terms of $\Delta$~Gyr 
per 0.1 dex) are similar for the two model sets. The individual values and the 
averages are listed in Table~\ref{alpha-fe}.

\subsection{\it Helium Abundance and Heavy Element Diffusion}

\begin{figure}
\epsscale{1.0}
\figurenum{14}
\plotone{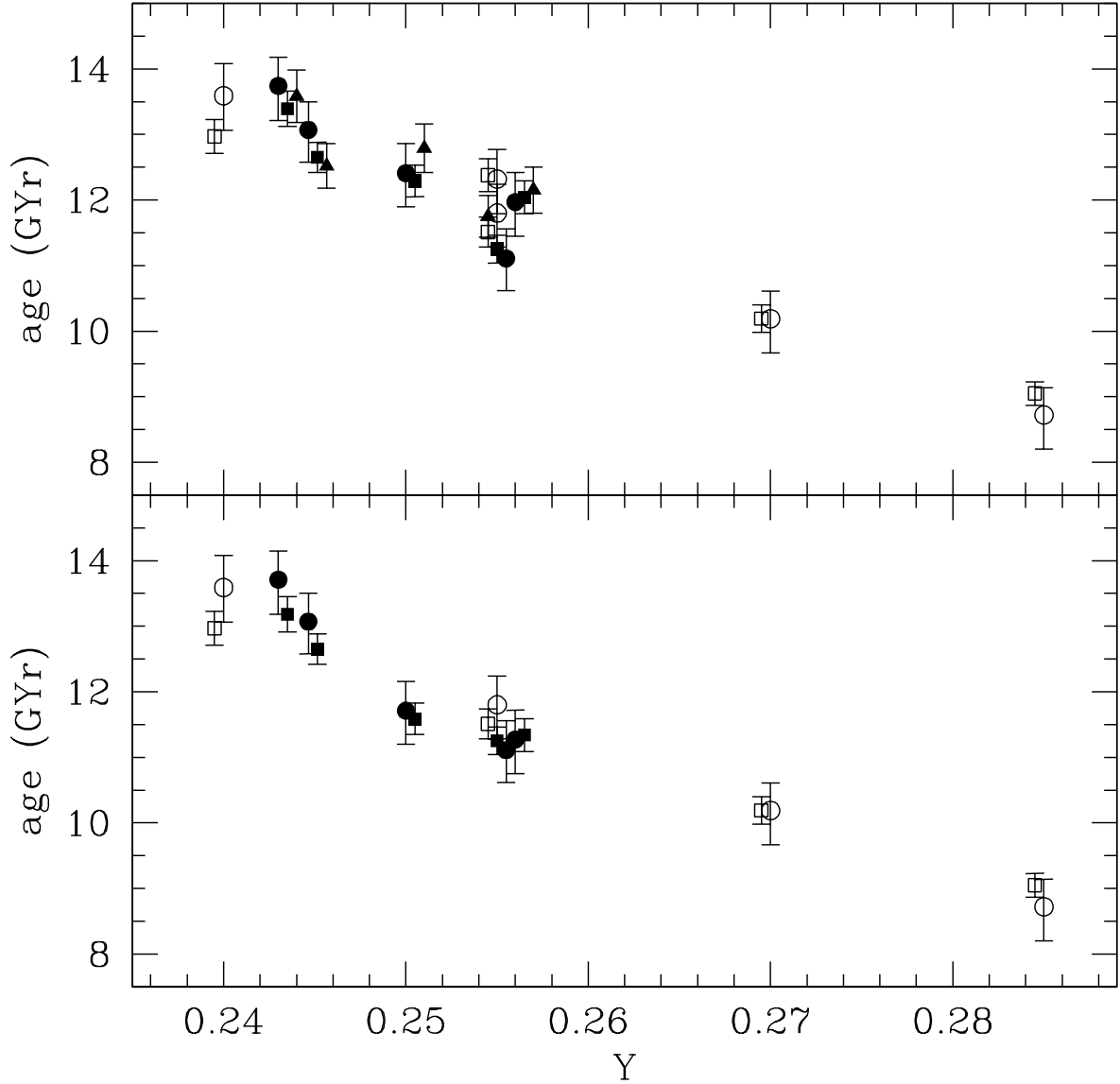}
\caption{
Age as a function of helium abundance as measured from Victoria-Regina, Yonsei-Yale, Padova,
Dartmouth, and Teramo isochrones (solid symbols plotted left to right, upper panel). 
Ages are determined from 
mass-luminosity relations (circles), mass-radius relations (squares),
and turnoff mass - age relations (triangles). Open symbols represent
ages measured from Dartmouth tracks, see text for details. Ages from two sets of Dartmouth
tracks are plotted for Y~=~0.255, models including helium and heavy element diffusion
(lower ages) and models without diffusion (higher ages). The plotted points have been
offset in helium abundance for clarity, the circles represent the correct
helium abundance for any one model set. Bottom panel: Ages corrected for 
the effects of diffusion (Victoria-Regina, Padova, and Teramo models). The
Victoria-Regina ages have been  additionally corrected to bring 
the $\alpha$-element enhancement to +0.4.
\label{fig14}}
\end{figure}

At the adopted [Fe/H], the models span a modest but important range in helium 
abundance ($Y$), from $Y$ = 0.243 for the Victoria-Regina models to 0.2555 for 
the Dartmouth models. The age estimates from mass-radius, mass-luminosity, and 
turnoff mass-age relations for all five model sets are plotted in the upper
panel of Fig.~\ref{fig14}. An obvious trend in age with $Y$ can be seen in 
Fig.~\ref{fig14} as expected. 
In order to further explore the dependence of age on $Y$, stellar evolution
tracks were computed by one of us (Dotter) using the code described by \citet{dotter}
for [Fe/H]~=~-0.71 and [$\alpha$/Fe]~=~+0.4 at $Y$= 0.24, 0.255, 0.27, 0.285, and 0.30 for the 
measured masses and $\pm$1-$\sigma$ uncertainties of the primary and secondary components of V69 
(see Table \ref{elements}). Ages were read from the tracks at the measured luminosities and 
radii of the components and plotted as open symbols in Fig.~\ref{fig14} for all but the 
$Y$=0.30 case.  The plotted track-based ages are a weighted average of 
the ages of the primary and  secondary components. These ages are listed in
Table~\ref{track-age}.
The variation in measured ages with helium abundance for these tracks closely 
follows that seen for the five sets of model isochrones.

Some of the scatter seen in the upper panel of Fig.~\ref{fig14} comes
from two sources. First, the Victoria-Regina models are calculated for
[$\alpha$/Fe]=+0.3, while the other four model sets use +0.4. To account for this 
we adjusted the Victoria-Regina models by +0.49~Gyr and +0.67~Gyr for ages determined 
from the mass-luminosity and mass-radius relations, respectively (see Fig.~\ref{fig13} 
and Table~\ref{alpha-fe}). Second, the Dartmouth and Yonsei-Yale models 
include the effects of helium and heavy element diffusion and gravitational settling
while the Padova, Teramo, and Victoria-Regina models do not.\footnote{In fact, 
the situation is more complicated than this because 
the Dartmouth models employ a treatment of diffusion that is inhibited in the 
outer 0.01 $M_\odot$ of the star \citep{chaboyer01}. The Teramo models 
employ diffusion in the calibration of their solar model but not in
models used to generate their metal poor models and thus their predicted ages 
are indirectly affected by diffusion.}
As a general rule, diffusion 
will lower the ages determined from models \citep{salaris00,chaboyer01}. To estimate the 
importance of diffusion in the measured ages, evolutionary tracks were also calculated
by Dotter for $Y = 0.255$ but without diffusion.  Ages determined from these tracks are 
plotted in the upper panel of Fig.~\ref{fig14}.  It might be expected that the effect 
of diffusion will depend on adopted helium abundance, and in the absence of a detailed 
comparison of diffusion in multiple model sets we adopt as a rough estimate a 
correction of -0.7~Gyr as measured from the Dartmouth tracks with and without diffusion. 
This correction was applied to the ages determined from the Padova, 
Teramo, and Victoria-Regina  models. We note that evidence for 
diffusion is seen in the globular cluster NGC~6397 \citep{korn07,lind08} based
on a systematic variation of measured metal abundances from the turnoff
through the giant branch.
No such evidence was detected  in 47~Tuc by \citet{koch}, although the
47~Tuc sample only included one turnoff star.

The corrected ages are plotted in the bottom panel of Fig.~\ref{fig14}. There are two 
conclusions to be drawn from this figure. First, the measured ages continue to be tightly 
correlated with $Y$, and in particular, remaining residuals from this trend are not strongly
related to [$\alpha$/Fe], distribution of the $\alpha$-element enhancements, 
or the mixing length. This correlation means that absolute ages of 
globular clusters based on individual stars in these clusters will be crucially dependent on 
an accurate measurement of the helium abundance of the cluster. This might seem impossible 
given the accuracy of various methods of determining cluster helium abundances  
\citep[see, for example,][]{sand00}.  However, we note that for  model sets
with $Y$~$\la$~0.25, the ages measured from mass-luminosity and mass-radius relations
differ systematically, with mass-luminosity relations predicting higher ages.
This effect is reversed for ages measured from Dartmouth tracks for $Y$~=~0.285.
If the models are sufficiently accurate, it should then be possible to constrain both age
and $Y$ by requiring that the mass-luminosity and mass-radius ages be consistent.

\begin{figure}
\epsscale{1.0}
\figurenum{15}
\plotone{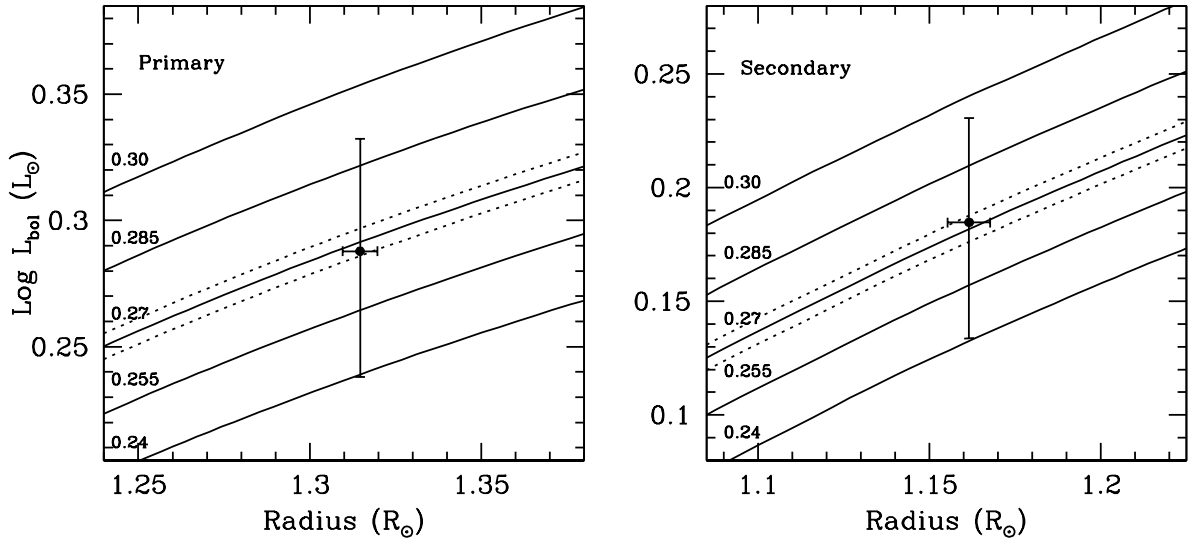}
\caption{
Dartmouth tracks calculated for the measured masses of the primary (left
panel) and secondary (right panel). In each case the tracks are calculated
for $Y$~=~0.24 through $Y$~=~0.30. The measured one sigma limits on mass
are plotted as dotted lines for the $Y$~=~0.27 tracks.
The measured values of radius and luminosity for the components of V69 are
both consistent with $Y$~=~0.27 with a one sigma range of 0.03.
\label{fig15}}
\end{figure}

We investigate the issue of helium abundance further in Fig.~\ref{fig15} where we 
plot the measured radii and luminosities of the components of V69 together with 
Dartmouth tracks calculated for the masses of the primary and secondary for a 
range values of the helium abundance. The measured  values of the radii and 
luminosities of both the primary and secondary agree to within the  uncertainties 
for $Y_p = 0.267^{+0.023}_{-0.029}$ and $Y_s = 0.271^{+0.024}_{-0.031}$  
with an  average value of $Y = 0.269^{+0.017}_{-0.021}$. This value can
be compared with estimates of the helium abundance estimated from the population
ratio $R = N_{HB}/R_{RGB}$ measured from
ground-based photometry ($Y = 0.216^{+0.013}_{-0.015}$: \citet{sand00}) and
HST photometry ($Y = 0.240\pm0.015$: \citet{salaris04}). The agreement is
reasonable given the uncertainties in the models used to calibrate the $R$
method. There are at least two possible ways to improve upon our estimate.
First, Figure 15 shows that our estimate is completely dominated by errors
in the luminosities of the components of V69. Near-infrared eclipse photometry
of V69 can be used to improve upon distenace estimates, and hence 
luminostiy estimates. Second, discovery of additional similar systems to
V69 in 47~Tuc will help improve improve upon distance, age, and 
helium abundance measurements for this cluster.

\citet{casagrande07} derived bolometric luminosities and effective
temperatures  from multi-band photometry and the infrared flux
method for a sample of 86 K-dwarfs. They compared these values to
Padova stellar isochrones to estimate the He to metal enrichment
ratio ($\Delta$Y/$\Delta$Z)
in the solar neighborhood.  Although their primary interest was to
measure $\Delta$Y/$\Delta$Z, Casagrande et al. found that, in an absolute
sense, the implied He content of the most metal-poor stars was as low as
Y$\sim$0.1--in marked disagreement with canonical values for the
primordial He content \citep{salaris04,spergel07}.  In contrast, we find
no indication that current stellar evolution models are unable to
reproduce the mass-radius relationship of V69.

\subsection{\it The Absolute Age of 47 Tuc}

As mentioned above, a measurement of the absolute age of 47~Tuc
will have random and systematic errors. Our best age estimates 
derived from the model isochrones have statistical errors of about
0.25 Gyr for mass-radius relations (see Table~\ref{ages}). Outside
of remaining systematic errors arising from different model 
parameters, the main systematic errors arise from the empirical
estimates for metallicity and $\alpha$-element enhancement. We
estimate the errors in both of these measurements to be 0.10~dex,
and following the results in Table~\ref{alpha-fe} we adopt a
systematic error of 0.85~Gyr, the quadratic sum of the contributions
from metallicity and $\alpha$-element enhancement. This leads to an absolute 
age estimate of 11.25$\pm0.21$(random)$\pm$0.85(systematic)~Gyr 
from Dartmouth isochrones adopting 
$Y$~=~0.255, [Fe/H]~=~-0.71, and [$\alpha$/H]~=~0.4. Note that the age
estimate from the Teramo isochrones is very similar when a correction
for diffusion has been applied (see Table~\ref{ages} and Fig~\ref{fig14}),
as are the ages derived from mass-luminosity relations for both the
Dartmouth and Teramo models. If we adopt a helium abundance of $Y$~=~0.27
then the Dartmouth tracks indicate an age of 10.21$\pm0.21$$\pm$0.85~Gyr.

This age estimate is consistent with  other recent estimates
of the age of 47~Tuc. For example,  \citet{gratton} find an age of 
$11.2\pm 1.1$ by fitting model isochrones with no diffusion to the 
cluster color-magnitude diagram, and an age of $10.8\pm 1.1$~Gyr for
models  including diffusion. \citet{sw02} derive an age
of $10.7\pm1.0$, also from isochrone fitting. \citet{zoccali} obtain an age of 
$13\pm 2.5$~Gyr by fitting the white-dwarf cooling sequence. 
\citet{grundahl02} concluded that the age of 47~Tuc  is "slightly below 12~Gyr"
based on analysis of Str\"{o}mgren photometry and isochrone fitting.
Our determination of the age of 47~Tuc based on 
observations of V69  has a  small statistical error 
compared to that derived from isochrone fitting.

\section{DISCUSSION AND SUMMARY}
\label{disc}
To the best of our knowledge, these measurements of the masses and
radii of the  components of V69 are the 
first such high accuracy (better than 1\%) measurements 
to be made for Population II stars. The binary is a member 
of the globular cluster 47~Tuc and so the determination of its 
distance and age applies to the cluster as well. 
We obtained a distance modulus of $(m-M)_{V}=13.35\pm 0.08$.
The main source of error in the distance estimate is the calibration dependent 
estimate of effective temperatures which we used to derive the bolometric
luminosities for the component stars.

A comparison of the measured masses, luminosities, and radii
of the components to stellar evolution models suggests that 
the age of the system and hence 
the globular cluster 47~Tuc can be measured to a statistical
accuracy of about 0.25 Gyr. However, it is important to 
understand the assumptions that go into any one model. In
particular, the derived ages are very sensitive to the adopted
helium abundance. We derive an age for 47~Tuc of 
11.25$\pm0.21$$\pm$0.45~Gyr for a helium abundance 
$Y$~=~0.255 using Dartmouth model isochrones. All models give similar ages 
when the effects of helium abundance are taken into account.

Comparison of Dartmouth evolutionary tracks calculated for the
measured masses of the primary and secondary indicate that the 
helium abundance can be measured to an accuracy of about 0.03 
for each of the components. We estimate a helium abundance of
$Y = 0.269^{+0.017}_{-0.021}$ for 47~Tuc. The measured masses, radii,
and luminosities of the components of V69 are consistent with Dartmouth models 
assuming [Fe/H] = -0.71, [$\alpha$/Fe] = + 0.4, and Y = 0.27.

The radii of
both stars are known with high accuracy, and it is therefore  possible
to obtain a more accurate and robust distance determination based
on the surface brightness method \citep{barnes76,lacy77,thompson01}. 
The empirical calibration of surface brightness relations for dwarf and subgiant
stars is improving
\citep{dibenedetto98,kervella,buermann06}, and it is reasonable to
imagine that a distance accurate to a few per cent can be measured with 
accurate radii and $(V-K)$ colors. We are in the process of collecting 
near IR eclipse profile photometry of both V69 and V228 \citep{kal07b} 
in order to measure the distances to these two binary stars in this way. 
These data will improve the estimates of the bolometric luminosities of 
the components and lead to a more accurate measurement of
the helium abundance and hence the absolute age of 47 Tuc.

Finally, we note that contributions to the errors in the radii are dominated 
by the photometric solution. Given the large inclination, the 
errors in the masses are completely dominated by the orbital solution. An
identical doubling of the existing set of radial velocity observations
leads to a 33$\%$ improvement in the mass estimates, and a subsequent improvement
in the age estimates. The system is relatively bright, and the prospects are good
that a substantial improvement in the measured masses can be achieved with further
radial velocity observations.

\acknowledgments

JK, WP and WK were supported by  grant N203~379936  
from the Ministry of Science and Higher Education, Poland.
Research of JK is also supported by the Foundation for Polish
Science through the grant MISTRZ.
IBT is supported by NSF grant AST-0507325.
Support from the Natural Sciences and Engineering Council of Canada
to SMR is acknowledged with gratitude. It is a pleasure to thank
John Southworth and Willy Torres for sharing software with us.  
Jay Anderson kindly measured the proper motion of V69. IBT acknowledges 
useful conversations with Chris Burns, Andy McWilliam, and George Preston.
This research used the facilities of the Canadian Astronomy Data Centre 
operated by the National Research Council of Canada with the support of 
the Canadian Space Agency. We dedicate this series of papers to the memory 
of our colleague Bohdan Paczy\'nski.

\begin{center}{\bf Appendix}
\end{center}

In their catalog of 47~Tuc variables  \citet{weldrake} included a potentially 
interesting detached binary named V39. The object has  an orbital period of 
4.6~d and is located about 0.30 mag to the red of the cluster
main-sequence on the $V/V-I$ plane. 
We have obtained $BV$ images of the V39 field using the 
du~Pont telescope. These images show that 
V39 is a close visual pair of stars separated by 1.5 arcsec.
The brighter component has $V\approx 18.1$ and $B-V\approx 0.99$ and is
located on a $V/B-V$ CMD among the Small Magellanic Cloud (SMC)
asymptotic giant  branch stars. The fainter component has
$V\approx 19.2$ and $B-V\approx 0.08$ and is a candidate SMC upper 
main sequence star. Given the orbital period of V39
we propose that it is the fainter component of the blend which
is the eclipsing binary.

\clearpage
\begin{deluxetable}{lccc}
\tablecolumns{6}
\tablewidth{0pt}
\tabletypesize{\normalsize}
\tablecaption{$BV$ Photometry of V69 at Minima and Quadrature
   \label{tab1}}
\tablehead{
\colhead{Phase}            &
\colhead{$V$}              &
\colhead{$B$}              &
\colhead{$B-V$}            
}
\startdata
Max & 16.836(1) & 17.384(2) & 0.548(2) \\
Min~I & 17.507(3) & 18.083(12) & 0.576(12) \\
Min~II & 17.461(3) & 18.011(14) & 0.550(14) \\
\enddata
\end{deluxetable}

\clearpage


\begin{deluxetable}{lrrc}
\tablecolumns{4}
\tablewidth{0pt}
\tabletypesize{\footnotesize}
\tablecaption{Radial Velocity Observations of V69 \label {rvdata}}
\tablehead{
\colhead{HJD}         &
\colhead{V$_{p}$}     &
\colhead{V$_{s}$}     &
\colhead{Phase}       \\
\colhead{(- 2400000)} &
\colhead{km s$^{-1}$} &
\colhead{km s$^{-1}$} &
\colhead{}       
}
\startdata  
53183.88089  & $-$57.33  &     24.10  & 0.173   \\
53201.93116  &    21.23  &  $-$55.92  & 0.784   \\
53206.81454  &  $-$5.58  &  $-$28.40  & 0.950   \\
53206.93425  &  $-$7.00  &  $-$27.15  & 0.954   \\
53210.82680  & $-$41.09  &      8.46  & 0.085   \\
53210.93896  & $-$42.27  &      9.98  & 0.089   \\
53271.78367  & $-$53.96  &     20.75  & 0.149   \\
53274.77303  & $-$59.04  &     26.63  & 0.250   \\
53280.65021  & $-$23.44  &   $-$9.68  & 0.449   \\
53281.65012  & $-$14.50  &  $-$18.34  & 0.483   \\
53282.69872  &  $-$7.02  &  $-$26.32  & 0.519   \\
53521.93280  &    12.54  &  $-$45.81  & 0.617   \\
53580.91364  &    11.43  &  $-$46.41  & 0.614   \\
53581.85305  &    16.04  &  $-$50.05  & 0.646   \\
53582.85552  &    19.69  &  $-$53.69  & 0.680   \\
53584.80006  &    22.23  &  $-$56.51  & 0.745   \\
53585.79822  &    21.34  &  $-$55.77  & 0.779   \\
53631.71181  & $-$50.28  &     16.17  & 0.334   \\
53633.74856  & $-$36.21  &      1.48  & 0.402   \\
53634.76492  & $-$27.12  &   $-$6.54  & 0.437   \\
53889.93336  & $-$37.97  &      6.09  & 0.075   \\
53891.93240  & $-$52.13  &     20.77  & 0.143   \\
53892.93208  & $-$56.74  &     24.68  & 0.177   \\
53893.93261  & $-$59.42  &     26.95  & 0.210   \\
53935.93349  &    14.75  &  $-$48.09  & 0.632   \\
53937.92995  &    21.18  &  $-$55.15  & 0.700   \\
53938.92894  &    22.39  &  $-$56.40  & 0.734   \\
\enddata
\end{deluxetable}

\clearpage

\begin{deluxetable}{lr}
\tablecolumns{2}
\tablewidth{0pt}
\tabletypesize{\normalsize}
\tablecaption{Orbital Parameters for V69
\label{orbit}}
\tablehead{
\colhead{Parameter}       &
\colhead{Value}
}

\startdata
$P$ (days)                       & 29.53975\tablenotemark{a} \\
$T_{0}$~(HJD-240 0000)           & 2453237.8421\tablenotemark{a} \\
$\gamma~(km~s^{-1})$             & -16.71$\pm$ 0.05 \\
$K_{p}~(km~s^{-1})$              & 41.03$\pm$0.13 \\
$K_{s}~(km~s^{-1})$              & 41.86$\pm$0.09 \\
$e$                              & 0.0563$\pm$0.0011 \\
$\omega$ (deg)                   & 149.15$\pm$1.86 \\
$\sigma_{p}~(km~s^{-1})$         & 0.50 \\
$\sigma_{s}~(km~s^{-1})$         & 0.33 \\
Derived quantities: & \\
$A~sin~i$~($R_{\odot}$)          & 48.301$\pm$0.095\\
$M_{p}~sin^{3}~i$~($M_{\odot}$)  & 0.8762$\pm$0.0048 \\
$M_{s}~sin^{3}~i~$($M_{\odot}$)  & 0.8588$\pm$0.0060 \\
\enddata

\tablenotetext{a}{Ephemeris adopted from photometry.}
\end{deluxetable}

\clearpage

\begin{deluxetable}{lr}
\tablecolumns{3}
\tablewidth{0pt}
\tabletypesize{\normalsize}
\tablecaption{Results of the Light Curve Analysis for V69
Obtained with the PHOEBE Code 
\label{lc2}}
\tablehead{
\colhead{Parameter}       & \\
}

\startdata
$i$ (deg)             & 89.771$\pm$ 0.009   \\
$T_{p}$~(K)  & 5945 \\
$T_{s}$~(K)  & 5959$\pm$3 \\
$e$                   & 0.0556$\pm$0.0002  \\
$\omega$~(deg)        & 149.72$\pm$0.25   \\
$r_{p}$               & 0.02727$\pm$0.00005 \\
$r_{s}$               & 0.02408$\pm$0.00008 \\
$(L_{s}/L_{p})_{V}$ &  0.7870$\pm$0.0012 \\
$(L_{s}/L_{p})_{B}$ &  0.7889$\pm$0.0015 \\
$\sigma_{rms}~(V)$~(mmag) & 10.0    \\
$\sigma_{rms}~(B)$~(mmag) & 14.5    \\
\enddata
\end{deluxetable}

\clearpage

\begin{deluxetable}{lrr}
\tablecolumns{3}
\tablewidth{0pt}
\tabletypesize{\normalsize}
\tablecaption{Results of the Light Curve Analysis for V69
Obtained with the JKTEBOP Code 
\label{lc1}}
\tablehead{
\colhead{Parameter}       &
\colhead{$V$} & \colhead{$B$}
}

\startdata
Adjusted quantities   &                      &     \\
$r_{p}+r_{s}$         & 0.05166$\pm$ 0.00009 & 0.05151$\pm$0.00021 \\
$k=r_{p}/r_{s}$       & 0.8836 $\pm$ 0.0072  & 0.8961$\pm$0.0199 \\
$i$ (deg)             & 89.768$\pm$ 0.012    & 89.769$\pm$0.028 \\
$J$                   & 1.0146$\pm$0.0031    & 1.0022$\pm$0.0091 \\
$e$                   & 0.0567$\pm$0.0008  & 0.0586$\pm$0.0025\\
$\omega$ (deg)        & 147.7$\pm$1.3       & 145.2$\pm$3.4\\
Other quantities      &                      &       \\
$r_{p}$               & 0.02722$\pm$0.00009  & 0.02717$\pm$0.00026\\
$r_{s}$               & 0.02405$\pm$0.00012  & 0.02435$\pm$0.00033\\
$L_{s}/L_{p}$         & 0.792$\pm$0.011    & 0.802$\pm$ 0.028 \\
$\sigma_{rms}$ (mmag) & 10.0                 & 14.5 \\
\enddata
\end{deluxetable}

\clearpage

\begin{deluxetable}{lr}
\tablecolumns{2}
\tablewidth{0pt}
\tabletypesize{\normalsize}
\tablecaption{Absolute Parameters for V69 
\label{elements}}
\tablehead{
\colhead{Parameter}       &
\colhead{Value}
}

\startdata
$A$~($R_{\odot}$)      & 48.30$\pm$0.12\\
$M_{p}$~($M_{\odot}$)  & 0.8762$\pm$0.0048 \\
$M_{s}$~($M_{\odot}$)  & 0.8588$\pm$0.0060 \\
$R_{p}$~($R_{\odot}$)  & 1.3148$\pm$0.0051 \\
$R_{s}$~($R_{\odot}$)  & 1.1616$\pm$0.0062 \\
$T_{p}$~(K)            & 5945$\pm$150 \\
$T_{s}$~(K)            & 5959$\pm$150 \\
$L^{bol}_{p}$($L_{\odot}$)&1.94 $\pm$0.21\\
$L^{bol}_{s}$($L_{\odot}$)&1.53 $\pm$0.17\\
$M_{\rm Vp}~(mag)$     &4.12 $\pm$0.11\\
$M_{\rm Vs}~(mag)$     &4.38 $\pm$0.12\\
${\rm log}~g_{p}({\rm cm~s^{-2}})$&4.143 $\pm$0.003\\
${\rm log}~g_{s}({\rm cm~s^{-2}})$&4.242 $\pm$0.003\\
\enddata

\end{deluxetable}

\clearpage

\begin{deluxetable}{llc}
\tablecolumns{3}
\tablewidth{0pt}
\tabletypesize{\normalsize}
\tablecaption{Comparison of Distance Determinations to 47~Tuc
\label{distance}}
\tablehead{
\colhead{Author}       &
\colhead{Method}       &
\colhead{$(m-M)_{V}$}
}

\startdata
\citet{gratton}    & Main Seq. fitting   & 13.50$\pm$0.08 \\
\citet{percival02} & Main Seq. fitting   & 13.37$\pm$0.11 \\
\citet{grundahl02} & Main Seq. fitting   & $13.33\pm 0.04\pm 0.1$ \\
\citet{zoccali}    & white dwarfs     & 13.27$\pm$0.14 \\ 
\citet{mclaughlin} & kinematics       & $13.15\pm 0.13$ \\
\citet{salaris07}  & IR luminosity of HB   & $13.18\pm 0.03 \pm 0.04$ \\
\citet{kal07b}     & Eclipsing binary & $13.40\pm 0.07$  \\
\citet{bono}       & Tip RG branch    & $13.32\pm 0.09$  \\
\citet{bono}       & IR lum. of RR Lyrae stars & $13.47\pm 0.11$ \\
This paper         & Eclipsing binary & $13.35\pm 0.08$ \\
\enddata

\end{deluxetable}

\clearpage

\begin{deluxetable}{lccccc}
\tablecolumns{5}
\tablewidth{0pt}
\tabletypesize{\normalsize}
\tablecaption{Parameters of Stellar Evolution Models
\label{models}}
\tablehead{
\colhead{Parameter}  &
\colhead{Dartmouth}  &
\colhead{Padova}     &
\colhead{Teramo}      &
\colhead{Victoria}   &
\colhead{Yonsei-Yale}      
}

\startdata
{[}Fe/H{]}   & $-$0.71  & -0.70 &   $-$0.70  & $-$0.705  &   $-$0.71  \\
{[$\alpha$}/Fe{]}   &  +0.40 & 0.40 &  +0.40  &  +0.30  &  +0.40  \\
$X$   &  0.738010  & 0.742000 &  0.736000  & 0.750250  &  0.748031  \\
$Y$   &  0.255000  & 0.250000 &  0.256000  & 0.243000  &  0.244646  \\
$Z$   &  0.006490  & 0.008000 &  0.008000  & 0.006750  &  0.007323  \\
{[}O/Fe{]} & 0.40  & 0.50 & 0.50 & 0.30 & 0.40  \\
mixing length &  1.938 & 1.680 & 1.913 & 1.890 & 1.743 \\
(Z/X)$_{solar}$ & 0.023 & 0.024 & 0.024 & 0.024 & 0.024 \\

\enddata
\end{deluxetable}

\clearpage


\begin{deluxetable}{llccccc}
\rotate
\tabletypesize{\scriptsize}
\tablecolumns{7}
\tablewidth{0pt}
\tabletypesize{\footnotesize}
\tablecaption{Ages in Gyr for the Components of V69 Derived from Model Isochrones
\label{ages}}
\tablehead{
\colhead{Method} &
\colhead{Component}       &
\colhead{Dartmouth}   &
\colhead{Padova}   &
\colhead{Teramo}   &
\colhead{Victoria-Regina} &
\colhead{Yonsei-Yale}   
}
\startdata
Mass - Luminosity & primary    & 11.11+0.54/-0.58 & 12.39+0.55/-0.62 & 11.93+0.55/-0.62 & 13.62+0.57/-0.66 & 13.02+0.54/-0.62 \\
                  & secondary  & 11.12+0.80/-0.93 & 12.46+0.80/-0.89 & 12.04+0.76/-0.93 & 13.90+0.68/-0.89 & 13.16+0.72/-0.81 \\
                  & average    & 11.11+0.45/-0.49 & 12.41+0.45/-0.51 & 11.97+0.45/-0.52 & 13.74+0.44/-0.53 & 13.07+0.43/-0.49 \\

\hline

Mass - Radius & primary    & 11.25$\pm0.26$ & 12.26$\pm0.29$ & 11.97$\pm0.31$ & 13.32$\pm0.33$ & 12.63$\pm0.29$ \\
              & secondary  & 11.25$\pm0.37$ & 12.34$\pm0.40$ & 12.16$\pm0.42$ & 13.54$\pm0.46$ & 12.70$\pm0.39$ \\
              & average    & 11.25$\pm0.21$ & 12.29$\pm0.24$ & 12.04$\pm0.25$ & 13.39$\pm0.27$ & 12.65$\pm0.23$ \\
\hline

Turnoff Mass & primary    & 10.91$\pm$0.24 & 11.81$\pm$0.27 & 11.19$\pm$0.25 & 12.50$\pm$0.29 & 11.57$\pm$0.26 \\
             & secondary  & 11.75$\pm$0.31 & 12.79$\pm$0.37 & 12.15$\pm$0.35 & 13.58$\pm$0.40 & 12.52$\pm$0.34 \\
\enddata

\clearpage

\end{deluxetable}

\begin{deluxetable}{lccc}
\tabletypesize{\normalsize}
\tablecolumns{4}
\tablewidth{0pt}
\tabletypesize{\footnotesize}
\tablecaption{Effect of [Fe/H] and [$\alpha$/Fe] on 
Measured Age\label{alpha-fe}} 
\tablehead{
\colhead{Method} &
\colhead{Model}       &
\colhead{[Fe/H]} &
\colhead{[$\alpha$/Fe]} \\
 & & $\Delta$Gyr~/~0.1~dex& $\Delta$Gyr~/~0.1~dex
}
\startdata
Mass - Luminosity & Dartmouth    & 1.09 & 0.67 \\
                  & Yonsei-Yale  & 1.03 & 0.67 \\
                  & average      & 1.06 & 0.67 \\

\hline

Mass - Radius & Dartmouth    & 0.73 & 0.44 \\
              & Yonsei-Yale  & 0.79 & 0.54 \\
              & average      & 0.76 & 0.49 \\
\hline

Turnoff Mass & Dartmouth    & 0.68 & 0.43 \\
             & Yonsei-Yale  & 0.78 & 0.37 \\
             & average      & 0.73 & 0.40 \\
\enddata

\end{deluxetable}

\clearpage

\begin{deluxetable}{llccccc}
\rotate
\tabletypesize{\scriptsize}
\tablecolumns{7}
\tablewidth{0pt}
\tabletypesize{\footnotesize}
\tablecaption{Ages in Gyr for the Components of V69 Derived from Dartmouth Tracks
\label{track-age}}
\tablehead{
\colhead{Method} &
\colhead{Component}       &
\colhead{Y = 0.24}   &
\colhead{Y = 0.255}   &
\colhead{Y = 0.255\tablenotemark{a}}   &
\colhead{Y = 0.27} &
\colhead{Y = 0.285}   
}
\startdata
Luminosity & primary    & 13.51+0.64/-0.66 & 11.77+0.56/-0.65 & 12.27+0.56/-0.65 & 10.20+0.52/-0.63 & 8.77+0.51/-0.62 \\
           & secondary  & 13.71+0.76/-0.87 & 11.85+0.72/-0.86 & 12.40+0.76/-0.93 & 10.16+0.71/-0.90 & 8.60+0.74/-0.93 \\
           & average    & 13.59+0.49/-0.53 & 11.80+0.44/-0.52 & 12.32+0.45/-0.53 & 10.19+0.42/-0.52 & 8.72+0.42/-0.52 \\

\hline

Radius & primary    & 12.97$\pm0.32$ & 11.53$\pm0.28$ & 12.32$\pm0.30$ & 10.23$\pm0.25$ & 9.10$\pm0.22$ \\
       & secondary  & 12.98$\pm0.45$ & 11.48$\pm0.40$ & 12.49$\pm0.43$ & 10.13$\pm0.36$ & 8.94$\pm0.32$ \\
       & average    & 12.97$\pm0.26$ & 11.51$\pm0.23$ & 12.38$\pm0.25$ & 10.21$\pm0.21$ & 9.05$\pm0.18$ \\

\enddata
\tablenotetext{a}{Tracks calculated without the effects of diffusion.}
\end{deluxetable}


\clearpage

\clearpage

\clearpage

\clearpage

\clearpage

\clearpage

\clearpage

\clearpage

\clearpage

\clearpage

\end{document}